\documentclass[twocolumn,dvipsnames]{aa} 

\usepackage{xcolor}
\usepackage{lipsum, babel}
\usepackage[varg]{txfonts}
\usepackage{rotating}
\usepackage{lineno}
\usepackage{caption}
\usepackage{graphicx}
\usepackage{amsmath}
\usepackage{lscape}
\usepackage{booktabs}
\usepackage[colorlinks=true,linkcolor=magenta,citecolor=blue,filecolor=black,urlcolor=cyan,final=true]{hyperref}

\newcommand\editone[1]{\textcolor{black}{#1}}

\graphicspath{{./}{figures/}}

\makeatletter
\renewcommand*\aa@pageof{, page \thepage{} of \pageref*{LastPage}}
\makeatother

\begin{document} 

   \title{Modeling CME encounters at Parker Solar Probe with OSPREI: Dependence on photospheric and coronal conditions}

   \author{
        Vincent~E.~Ledvina\inst{1}\thanks{now at Geophysical Institute, University of Alaska Fairbanks, Fairbanks, AK 9709, USA}
        \and
        Erika~Palmerio\inst{1}
        \and
        Christina~Kay\inst{2,3}
        \and
        Nada~Al-Haddad\inst{4}
        \and
        Pete~Riley\inst{1}
        }

   \institute{
        Predictive Science Inc., San Diego, CA 92121, USA\\
        \email{vledvina@alaska.edu}
        \and
        Heliophysics Science Division, NASA Goddard Space Flight Center, Greenbelt, MD 20771, USA
        \and
        Department of Physics, The Catholic University of America, Washington, DC 20064, USA
        \and
        Space Science Center, University of New Hampshire, Durham, NH 03824, USA
        }
   
   \titlerunning{Modeling CMEs at PSP with OSPREI}
   \authorrunning{Ledvina et al.}

   \date{Received; accepted}

  \abstract
  {Coronal mass ejections (CMEs) are eruptions of plasma from the Sun that travel through interplanetary space and may encounter Earth.  CMEs often enclose a magnetic flux rope (MFR), the orientation of which largely determines the CME's geoeffectiveness.  Current operational CME models do not model MFRs, but a number of research ones do, including the Open Solar Physics Rapid Ensemble Information (OSPREI) model.}
  {We report the sensitivity of OSPREI to a range of user-selected photospheric and coronal conditions.}
  {We model four separate CMEs observed in situ by Parker Solar Probe (PSP).  We vary the input photospheric conditions using four input magnetograms (HMI Synchronic, HMI Synoptic, GONG Synoptic Zero-Point Corrected, and GONG ADAPT).  To vary the coronal field reconstruction, we employ the Potential-Field Source-Surface (PFSS) model and we vary its source-surface height in the range 1.5--3.0\,R$_{\odot}$ with 0.1\,R$_{\odot}$ increments.}
  {We find that both the input magnetogram and PFSS source surface often affect the evolution of the CME as it propagates through the Sun's corona into interplanetary space, and therefore the accuracy of the MFR prediction compared to in-situ data at PSP. There is no obvious best combination of input magnetogram and PFSS source surface height.}
  {The OSPREI model is moderately sensitive to the input photospheric and coronal conditions.  Based on where the source region of the CME is located on the Sun, there may be best practices when selecting an input magnetogram to use.}


\maketitle


\newpage
\section{Introduction} \label{sec:intro}

A coronal mass ejection (CME) is an eruption on the Sun that releases significant amounts of plasma into interplanetary space \citep[see more in][]{webb2012}. As plasma moves out from the Sun, the magnetic field becomes ‘frozen-in’ so that it remains fixed within the plasma parcels in which it is embedded close to the Sun. Due to this frozen-in flux property of space plasmas, CMEs carry with them an embedded magnetic field in the form of a flux rope \citep[e.g.,][]{green2018}.  Occasionally, CMEs launched from the Sun will be Earth-directed, and when they come in contact with Earth’s magnetic environment, they may cause geomagnetic storms that damage power grids, cause spacecraft charging, and create auroral displays \citep[e.g.,][]{pulkkinen2007}.  The strength and type of interaction a CME produces has much to do with its speed, ram pressure, and magnetic configuration.  Current solar wind models used by space weather forecasting offices to model CMEs can only predict their speeds, densities, and arrival times operationally \citep{pizzo2011}.  Other CME models that are presently used in research use alternative workflows to predict not only speed, density, and arrival time, but also the configuration of the magnetic flux rope (MFR) embedded inside the CME. Magnetohydrodynamic (MHD) models are generally more realistic, but also complex and computationally-expensive, and these include the Magnetohydrodynamic Algorithm outside a Sphere \citep[MAS;][]{mikic1999} code coupled with the modified Titov--D{\'e}moulin flux rope \citep{torok2018}, the Alfv{\'e}n Wave Solar Model \citep[AWSoM;][]{oran2013} coupled with the Gibson--Low flux rope \citep{jin2017}, and the European Heliospheric Forecasting Information Asset \citep[EUHFORIA;][]{pomoell2018} coupled with the Spheromak \citep{verbeke2019} or the FRi3D \citep{maharana2022} flux rope models. On the other hand, analytical models are generally less realistic, but at the same time simpler and computationally efficient. Such models include the Three‐Dimensional Coronal ROpe Ejection \citep[3DCORE;][]{mostl2018}, the INterplanetary Flux ROpe Simulator \citep[INFROS;][]{sarkar2020}, and the Open Solar Physics Rapid Ensemble Information \citep[OSPREI;][]{kay2022a}. Most notably, analytical models can achieve MFR predictions quickly with relatively easily-obtained input parameters. 

The orientation of the MFR inside a CME is correlated with how much energy is transferred from the solar wind into Earth’s magnetic system and has a great effect on what impacts are seen on Earth’s surface and in the near-Earth space environment.  An MFR with a long-duration negative B$\mathrm{_{z}}$ component is required for strong geomagnetic storms, as it opens the subsolar magnetopause via magnetic reconnection, allowing for the transfer of energy, plasma, and momentum from the solar wind into Earth’s magnetosphere \citep[e.g.,][]{dungey1961}.  Predictions of an MFR’s B$\mathrm{_{z}}$ component are important for understanding its impact at Earth, but current forecasts are still unable to model CME MFRs, resulting in frequent over and under-estimations of CME geoeffectiveness.  For example, the strength of the “St Patrick’s Day Storm” on 17 March 2015 \citep{kataoka2015,wu2016} was underestimated by the National Oceanic and Atmospheric Administration’s Space Weather Prediction Center (NOAA/SWPC), which predicted only maximum G1 geomagnetic storm conditions.  The CME shock arrived 15 hours ahead of forecasts and the MFR contained a large southward B$\mathrm{_{z}}$ component, producing a G4-level geomagnetic storm.  Conversely, one of the first notable space weather events of solar cycle 25, a partial halo CME eruption on 28 March 2022, triggered a G3 geomagnetic storm watch from NOAA SWPC for 31 March 2022.  While the arrival time of the CME was accurately forecasted, due to an MFR with a strong positive B$\mathrm{_{z}}$ configuration, the CME only managed to create a brief G1-level geomagnetic storm. These are only two examples of where best-effort space weather forecasting simply does not do enough to account for the contribution of a CME's MFR.

Deflections of the CME propagating through the Sun's corona may also affect the orientation of the MFR and CME's speed and trajectory, affecting the accuracy of the forecast arrival time.  In one example, \citet{mays2015} and \citet{mostl2015} showed that the major CME of 7 January 2014 was not forecast accurately due to rotations and deflections before propagating in the solar wind. Improving the orientation and directionality of the CME through more advanced modeling improved the forecast arrival time by over 18 hours at Earth.
While there are a host of models that can model MFRs and coronal deflections of CMEs, for them to be operationally viable, they need to be validated and their sensitivities to different input parameters assessed.  

In this work, we investigate the OSPREI model and determine its sensitivity to varying input photospheric and coronal conditions. We compare OSPREI model results with Parker Solar Probe \citep[PSP;][]{fox2016} in-situ solar wind measurements.  PSP is used as a ground truth instead of Earth so that we can contextualize our results to space weather predictions at any point in the heliosphere assuming that maps of the solar photospheric field are only captured from Earth's viewpoint. \editone{At the times of the CME events analyzed here, PSP was located at heliocentric distances between ${\sim}$0.45 and ${\sim}$0.75~au, so one may expect forecasting uncertainties to decrease as the spacecraft gets closer to the Sun. Nevertheless, in our investigation we do not focus on CME arrival time, but rather on the MFR magnetic structure. Although it has been shown that CMEs can deflect and rotate in interplanetary space \citep[e.g.,][]{isavnin2014}, it is well known that most changes in trajectory and orientation take place in the solar corona \citep[e.g.,][]{kay2015b}. Hence, we can assume the impact of these contributions to decrease with radial distance so rapidly to not significantly affect our results.} The findings \editone{of this work} will be useful for incorporating OSPREI as a potential future operational CME model. 


The outline of this manuscript is as follows:
We begin with an overview of OSPREI and by detailing the modeling setup used for our analysis in Sect.~\ref{sec:setup}. In Sect.~\ref{sec:data_analysis}, we discuss the data sets used and our methodology to vary the photospheric and coronal conditions as inputs into the OSPREI model.  In Sect.~\ref{sec:results} we describe the model's reaction to our varying input parameters on the four CMEs, and we discuss these results in Sect.~\ref{sec:discuss}. Finally, in Sect.~\ref{sec:conclude} we interpret our findings in the context of other CME models and make suggestions as to how OSPREI may be used in real-time forecasting situations.


\section{Modeling Setup} \label{sec:setup}

In this section, we describe  the modeling setup used in this study: We provide a brief overview of the OSPREI modeling suite (Sect.~\ref{subsec:osprei}), and then follow with a description of the different photospheric (Sect.~\ref{subsec:magnetograms}) and coronal (Sect.~\ref{subsec:pfss}) inputs that we explore to test the model's sensitivity.


\subsection{Overview of OSPREI} \label{subsec:osprei}

OSPREI models Sun-to-Earth---or, more generally, Sun-to-heliosphere---CME behavior beyond more traditional interplanetary propagation models, including internal thermal and magnetic field properties of the CME.  OSPREI has three components: (1) the Forecasting a CME’s Altered Trajectory \citep[ForeCAT;][]{kay2015a} module calculates the CME's deflections and rotations as it propagates through the solar corona; (2) the Another Type of Ensemble Arrival Time Results \citep[ANTEATR;][]{kay2018} module calculates the heliospheric propagation and arrival time of the CME at a given point in interplanetary space; and (3) the ForeCAT In situ Data Observer \citep[FIDO;][]{kay2017} models the CME flux rope's magnetic field as a time series at the location of interest. \editone{We remark that all CME deflections and rotations take place within the ForeCAT (coronal) domain, while in the ANTEATER (interplanetary) one CMEs are assumed to maintain their trajectory.} For more information on these modules and OSPREI's capabilities, see \citet{kay2022a}.

In this work, \editone{we set the outer boundary of the ForeCAT (coronal) domain to 20~R$_{\odot}$. Additionally,} we use the Physics-driven Approach to Realistic Axis Deformation and Expansion version of ANTEATR \citep[ANTEATR-PARADE;][]{kay2021a}, which includes the Elliptic-Cylindrical (EC) analytical flux rope model of \citet{nieves-chinchilla2018} and is able to model physics-driven changes in the size and shape of the CME's central axis as well as cross-section during propagation. We note that the sensitivity of ANTEATR-PARADE---and, as a consequence, of OSPREI---to CME input parameters has been explored in \citet{kay2021b}. Here, we focus instead on OSPREI's sensitivity to the initial photospheric and coronal conditions, while maintaining the CME input parameters fixed.


\subsection{Magnetogram Sources} \label{subsec:magnetograms}
OSPREI’s FORECAT module requires realistic modeling of the global solar corona to compute deflections and rotations of the CME flux rope as it propagates through the corona. Since direct measurements of the coronal magnetic field are not performed routinely, models are usually employed to extrapolate the global configuration of the solar corona using photospheric field measurements as boundary condition \citep[e.g.,][]{wiegelmann2017}. Hence, the first source of uncertainty when modeling CME propagation through the solar corona concerns the chosen input photospheric field map. Several studies have tested the influence of input magnetograms on MHD modeling results of the ambient solar wind and its transients \citep[e.g.,][]{riley2021,jin2022} but, to our knowledge, similar studies have not been performed in the context of analytical CME modeling. In this section, we present the four types of magnetograms that we use in this work to test OSPREI's sensitivity to user-selected input conditions: HMI Synchronic, HMI Synoptic, GONG Zero-Point Corrected, and GONG ADAPT \#10. We note that, in previous applications, OSPREI has only employed the HMI Synchronic or Synoptic magnetograms in its ForeCAT module.

\subsubsection{HMI Synoptic} \label{subsubsec:synoptic}
The Helioseismic and Magnetic Imager \citep[HMI;][]{scherrer2012} on board the Solar Dynamics Observatory \citep[SDO;][]{pesnell2012} provides space-based, full-disk photospheric vector magnetic field measurements with a 12-minute cadence.  We use synoptic maps made from the imputed radial component of the magnetic field over the entire solar disk. Synoptic maps are generated using 20 magnetograms captured close to the time of the central meridian passage (CMP) for that longitude, so the effective temporal width of the HMI synoptic map contribution is about three hours.  In general, these individual magnetograms are all captured within ${\sim}2^{\circ}$ of the CMP.  After a full Carrington rotation, a full synoptic map is created representing photospheric magnetic fields at all Carrington longitudes.  Because the standard synoptic map is constructed from data observed within $2^{\circ}$ of the central meridian, each longitude is both a different physical location and the field measured at a different time.  While central meridian observations provide the best observational quality and radial magnetic field calculations, synoptic maps by nature poorly represent fast-evolving features such as active regions.

\subsubsection{HMI Synchronic} \label{subsubsec:synchronic}
HMI synchronic maps involve the same instruments, observational methods, and image processing routines as HMI synoptic maps, but are designed to provide a full-Sun snapshot that better represents the magnetic conditions on the Earth-facing solar disk. Synchronic (daily) maps replace a $120^{\circ}$ longitude range of data centered around the central meridian from the original synoptic map with data observed at a one synchronized time \citep[see, e.g.,][]{hayashi2015}. 

\subsubsection{GONG Zero-point Corrected} \label{subsubsec:gongzero}
The Global Oscillations Network Group \citep[GONG;][]{harvey1996} is a program operated by the National Solar Observatory, whose six telescopes provide 24-hour coverage, low-noise, near-real-time, precise synoptic maps of the photospheric magnetic field  \citep{hill2018}.  The reliability, high cadence, and spatial coverage of GONG magnetograms make them appealing for operationally-oriented space weather applications, including the WSA--Enlil model employed at NOAA/SWPC \citep{pizzo2011, steenburgh2014} and the EUHFORIA model employed at the ESA Virtual Space Weather Modelling Centre \citep{poedts2020}.  One factor impacting the quality of the magnetogram observations from the GONG network stems from non-uniformities and small imperfections in each observatory's magnetogram modulator.  These can introduce uncertainties of several Gauss in the zero point of the magnetograms, and these uncertainties can be amplified when data are merged from multiple GONG sites.  Areas of quiet Sun are most affected by these zero-point errors since the surface magnetism is sometimes on the order of tens of Gauss \citep[e.g.,][]{bellotrubio2019}.  Thus, the current method of zero-point correction involves fitting a low-order 2D polynomial surface to quiet-Sun regions, with careful attention to exclude any contribution from active regions.  The resulting fitted surface is used as the zero point.  These zero-point corrected synoptic magnetograms are important for extrapolations of the solar corona using models since open magnetic field lines are often rooted in regions of weaker magnetic field.

\subsubsection{GONG ADAPT Realization \#10} \label{subsubsec:gongadapt10}
The Air Force Data Assimilative Photospheric Flux Transport \citep[ADAPT;][]{arge2010, arge2010_2, arge2013} model is a modified version of the \citet{worden&harvey2000} model and accounts for differential rotation, meridional flow, supergranular diffusion, and random flux emergence, thus, creating more accurate estimates of magnetic flux distribution on the solar surface.  GONG ADAPT magnetograms combine photospheric magnetic field observations from the GONG telescope network with the ADAPT flux transport model, producing 12 possible realizations of the photospheric magnetic field.  The ADAPT realizations are synchronic maps. For this study, we randomly select the ADAPT realization \#10.


\subsection{Coronal Extrapolations} \label{subsec:pfss}

As mentioned in the previous section, models are needed to extrapolate the coronal magnetic field from photospheric maps. Hence, the second source of uncertainty when modeling CME propagation through the solar corona is related to the coronal model assumptions and choice of input parameters. In this work, we employ the widely-used Potential Field Source Surface \citep[PFSS;][]{wang1992} model, which neglects electric currents in the corona and represents the global field using a scalar potential. Magnetic forces dominate the low corona (i.e., the plasma beta is low), but at larger heights the weaker fields are dragged out with the solar wind, becoming essentially radial. To simulate this effect, the PFSS model sets the scalar potential to be constant on a certain surface---the source surface \citep{schatten1969}. Above the PFSS source surface, the coronal field lines are forced to be radial, modeling the effect on the field of the outflowing solar wind.

Historically and in most applications to this day, the PFSS source surface has been set to be at R$_\mathrm{ss}$ = 2.5\,R$_{\odot}$ \citep{altschuler1969}---and this is also valid for previous simulations ran with OSPREI. The location of the heliospheric current sheet (HCS) resulting from setting the PFSS source surface at 2.5\,R$_{\odot}$ is shown for each of the input magnetograms in Figs.~\ref{fig:magnetograms_event1}--\ref{fig:magnetograms_event4}. Nevertheless, a number of studies have questioned the use of a single, fixed value for R$_\mathrm{ss}$, suggesting e.g.\ that it should be lowered to better represent the interplanetary magnetic field during solar minimum \citep{lee2011}, to reconstruct the areas of coronal holes \citep{asvestari2019}, or to (at least partially) resolve the open flux problem \citep{riley2019}. \citet{arden2014} proposed the use of a ``breathing'' source surface, with its height changing with the phase of the solar cycle. Furthermore, the spherical shape of the PFSS source surface is even questioned by some studies \citep[e.g.,][]{schulz1978, schulz1997, levine1982, riley2006, kruse2020}. These works and others have tested R$_\mathrm{ss}$ values approximately within the range 1.2--3.5\,R$_{\odot}$. In our study, we vary R$_\mathrm{ss}$ in the range 1.5--3.0\,R$_{\odot}$ with 0.1\,R$_{\odot}$ increments, i.e., we associate each of the four magnetograms presented in Sect.~\ref{subsec:magnetograms} with 16 different PFSS extrapolations, yielding a total of 64 combinations of photospheric and coronal conditions to test for each event.


\section{Event Selection and Analysis} \label{sec:data_analysis}

In this section, we first provide an overview of the four CME events observed in situ by PSP that we use to investigate the effect of input magnetogram and PFSS source surface on OSPREI results (Sect.~\ref{subsec:events}). Then, we describe our analysis of the input magnetograms used for each event (Sect.~\ref{subsec:maginputs}) and the OSPREI input parameters that we set for each CME (Sect.~\ref{subsec:inputs}).


\subsection{Events} \label{subsec:events}

\begin{figure*}[t!]
 \resizebox{0.8\hsize}{!}{\includegraphics[angle=0]{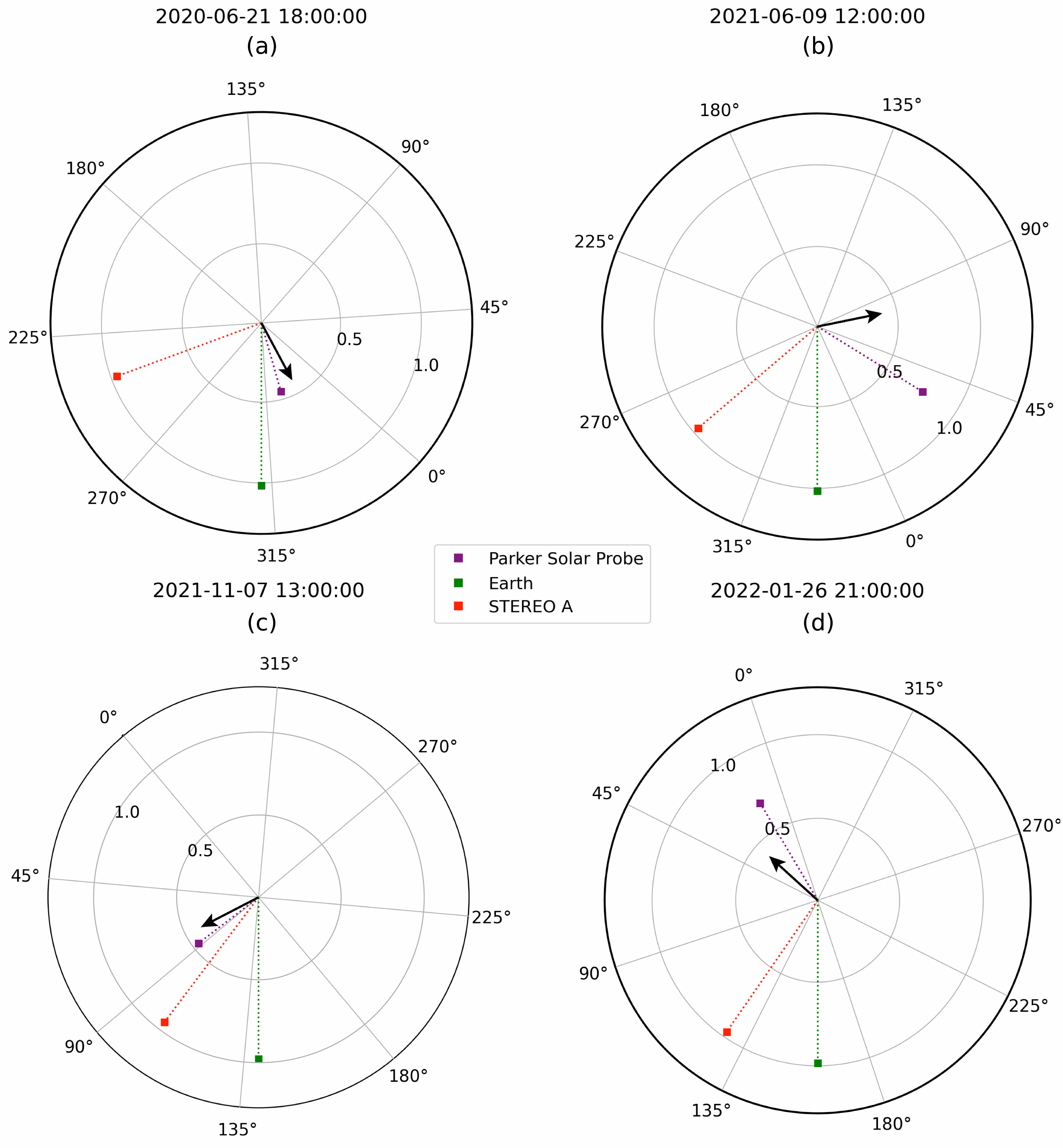}}
  \centering 
  \caption{Positions of PSP, Earth, and STEREO-A at the CME onset time for (a) Event~1, (b) Event~2, (c) Event~3, and (d) Event~4. The arrow in each panel indicates the source longitude of the corresponding CME. The longitudes shown are in Carrington coordinates. The plots are made with the Solar MAgnetic Connection HAUS \citep[Solar-MACH;][]{gieseler2023} tool.}
  \label{fig:orbits}
  \end{figure*}

In this parameter study, we model four CME events, with source regions placed at very different locations with respect to Earth's viewpoint (see Fig.~\ref{fig:orbits}).  The first CME event is a streamer blowout on 21 June 2020 that erupted from close to the central meridian.  The second CME event is a northwestern limb eruption on 9 June 2021 originating from an active region just off the visible solar disk.  The third CME event occurred on 7 November 2021 from an active region close to the northeastern limb of the Sun.  Finally, the fourth CME event occurred on 26 January 2022 from an active region on the far side of the Sun. For each event, we provide an overview of the available remote-sensing observations as Supplementary Movies. These observations come from two viewpoints, namely Earth and the Solar Terrestrial Relations Observatory Ahead \citep[STEREO-A;][]{kaiser2008} spacecraft, orbiting the Sun from a heliocentric distance of ${\sim}$1~au---with variable longitudinal separation with respect to Earth. Earth-based observations come from the Atmospheric Imaging Assembly \citep[AIA;][]{lemen2012} on board SDO, imaging the solar disk, as well as the C2 and C3 cameras part of the Large Angle and Spectrometric Coronagraph \citep[LASCO;][]{brueckner1995} on board the Solar and Heliospheric Observatory \citep[SOHO;][]{domingo1995}, imaging the solar corona. From STEREO-A, we use data from the Extreme Ultraviolet Imager (EUVI) solar disk telescope and the COR2 coronagraph, both part of the Sun-Earth Connection Coronal and Heliospheric Investigation \citep[SECCHI;][]{howard2008} suite.

\subsubsection{Event 1}
The first CME that we analyze in this work is a streamer blowout initiated on 21 June 2020 around 18:00~UT from close to the central meridian of the Earth-facing Sun (Fig.~\ref{fig:orbits}(a)). For full details related to this event, please see \citet{palmerio2021b} and \citet{pal2022}. An overview of the remote-sensing observations associated with this event is shown in Movie~1. Notably, this type of eruption is what is known as a ``stealth CME.'' A CME is stealth if no distinct low-coronal signatures \citep[such as coronal dimmings, coronal waves, filament eruptions, flares, and post-eruptive arcades;][]{hudson2001} can be found on the solar disk---a ``CME from nowhere'' \citep{robbrecht2009,palmerio2021a}.  Thus, the magnetic fields of stealth CMEs are particularly hard to model and forecast since it is often difficult to determine a well-defined source region and flux rope configuration.  Since stealth CMEs can still drive major geomagnetic disturbances at Earth \citep{nitta2017,nitta2021}, it bears merit to analyze this particular event so that it may lead to better techniques for modeling stealth CMEs in the heliosphere.

\subsubsection{Event 2}
The second CME that we analyze in this work occurred on 9 June 2021 around 12:00~UT. See Movie~2 for an overview of remote-sensing observations of this event.  Starting around 12:00 UT, a large eruption was seen by SDO/AIA.  The source of the eruption was an active region off the northwestern limb of the solar disk ($\sim$23.5$^{\circ}$ latitude and $\sim$77$^{\circ}$ longitude in Carrington coordinates, see Fig.~\ref{fig:orbits}(b)).  The eruption evacuated the corona, generating a blast wave that propagated eastward from the limb across the solar disk.  A flux rope structure was also seen leaving the corona off the Sun's limb behind the initial bright front of the CME.  We examine coronagraph images from the SOHO/LASCO C2 and C3 telescopes and find a well-defined three-part structure \citep[i.e., consisting of a bright front, a cavity, and a core;][]{illing1985,vourlidas2013} with a flux rope visible in the darker region as the CME propagated out of the C3 field of view. This event was selected because at the eastern and western limbs, ``stitching'' occurs in synchronic magnetogram maps that may create different coronal field reconstructions when compared to diachronic magnetogram maps, thereby altering the OSPREI model outputs.

\subsubsection{Event 3}
The third CME that we analyze in this work occurred on 7 November 2021 around 13:00~UT.  See Movie~3 for an overview of remote-sensing observations of this event.  The source of the eruption was an active region on the northeastern limb of the Sun as seen from SDO (Fig.~\ref{fig:orbits}(c)), with a position in Carrington coordinates of $\sim$15$^{\circ}$ latitude and $\sim$78$^{\circ}$ longitude.  As seen in Movie~3, the source region of the event was also seen by STEREO/SECCHI/EUVI-A.  In coronagraph imagery, the CME is visible in the C2 and C3 telescopes on board SOHO's LASCO as a faint white-light plume extending away from the left-hand side of the occulting disk.  The STEREO-A spacecraft was nearly in quadrature ($\sim$90$^{\circ}$ east of the Sun-Earth line) with Earth, and therefore in the COR2 coronagraph on board SECCHI the CME was observed as a faint full-halo signature, better identified in base-difference images \citep[e.g.,][]{attrill2010}.  This event was selected because it represents an almost base-case scenario for OSPREI: An eruption on the Earth-facing disk is optimal for magnetogram reconstruction of the source region, and multiple space-based coronagraph and low-coronal observations make other CME parameters (tilt, speed, chirality, etc.) easier to determine. Moreover, this event originated from ${\sim}60^{\circ}$ east of the central meridian of the Earth-facing Sun, i.e., in the vicinity of where the ``stitching'' occurs in synchronic magnetogram maps.

\subsubsection{Event 4}
The fourth and final CME that we analyze in this work is an eruption that took place on 26 January 2022 around 21:00~UT.  See Movie~4 for an overview of remote-sensing observations of this event.  The eruption initiated from an active region on the Sun's far side.  SDO imagery shows no evidence of the CME using AIA, but the EUVI instrument on board STEREO-A did capture post-eruptive arcades \citep[e.g.,][]{tripathi2004} from an active region approaching the eastern limb of the visible disk.  Based on magnetograms rendered at the time of the eruption showing an active region in this general location, we determine that the CME initiated from an active region at $\sim$~-17$^{\circ}$ latitude and $\sim$~30$^{\circ}$ longitude in Carrington coordinates (Fig.~\ref{fig:orbits}(d)). STEREO/SECCHI/COR2-A observed a faint CME, while in SOHO/LASCO the eruption is more visible.  This event was chosen because it can be used to test the differences in model outputs for a far-sided event, where no ``recent'' magnetograph observations are available, while the limb measurements from STEREO-A give us confidence on the location from where the CME launched.


\subsection{Magnetogram Analysis} \label{subsec:maginputs}

 \begin{figure*}[t!]
 \resizebox{1\hsize}{!}{\includegraphics[angle=0]{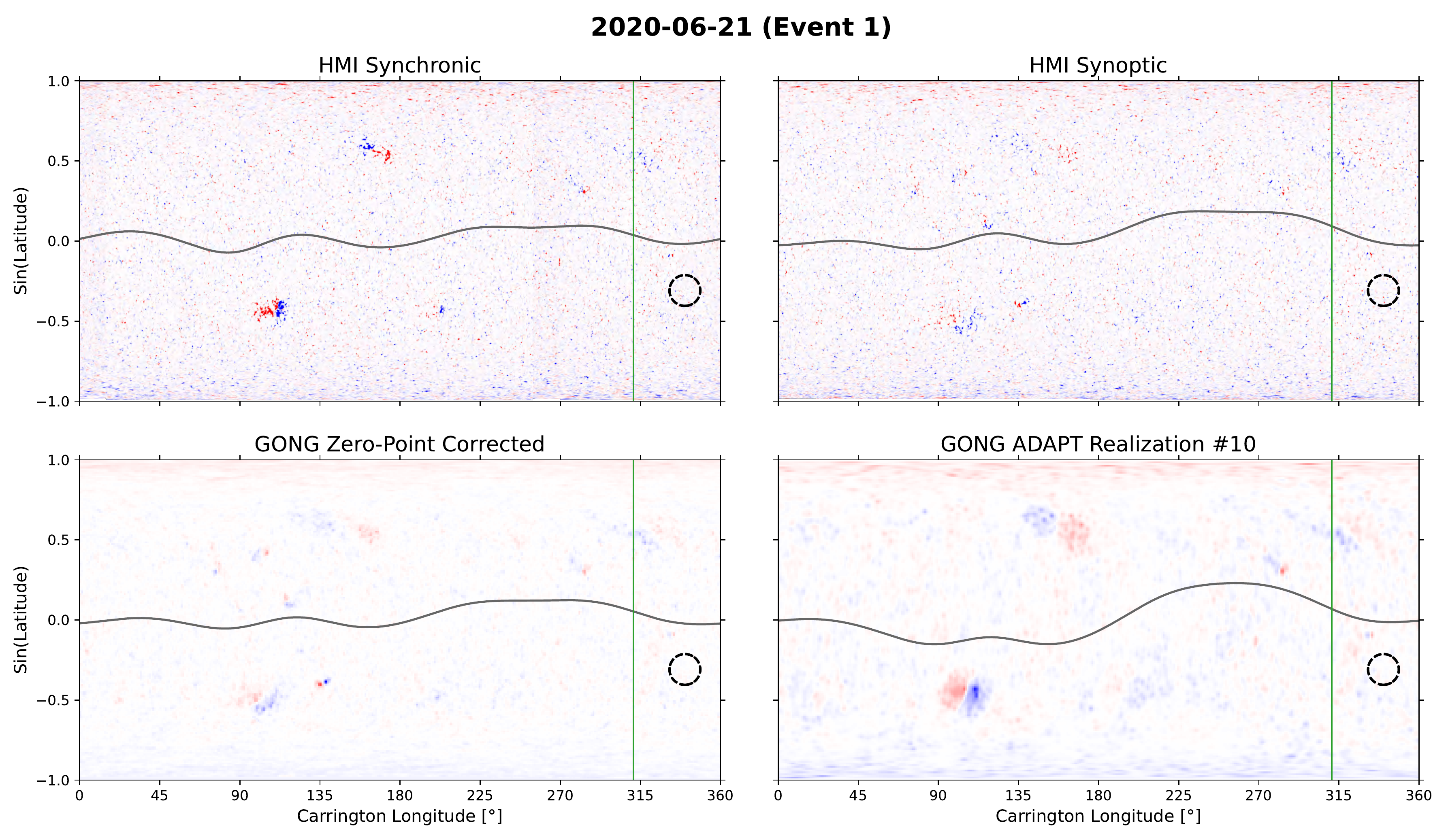}}
  \centering 
  \caption{The four magnetograms that we use as input photospheric conditions for Event~1.  \textit{Top left}: HMI Synchronic; \textit{Top right}: HMI Synoptic; \textit{Bottom left}: GONG Zero-point corrected; \textit{Bottom right}: GONG ADAPT Realization \#10. The green vertical line indicates the Carrington longitude of the central meridian as seen from Earth. The dotted circle marks the approximate location of the CME source region. The gray curve marks the location of the heliospheric current sheet at 2.5~R$_\odot$. All maps are saturated to ${\pm}100$~G, with red (blue) indicating positive (negative) polarity.}
  \label{fig:magnetograms_event1}
  \end{figure*}
  
\begin{figure*}[t!]
\resizebox{1\hsize}{!}{\includegraphics[angle=0]{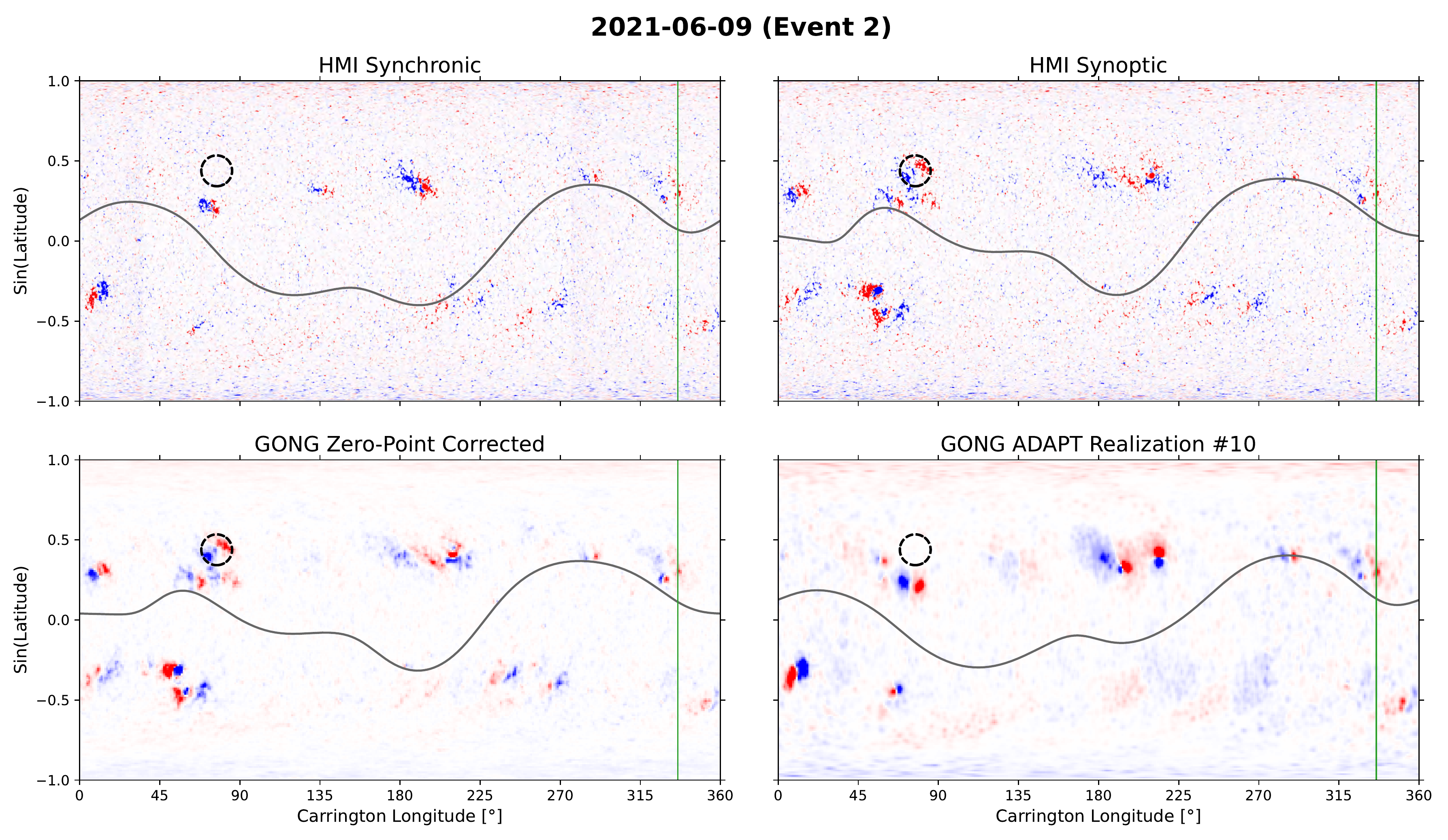}}
 \centering 
 \caption{The four magnetograms that we use as input photospheric conditions for Event~2.  \textit{Top left}: HMI Synchronic; \textit{Top right}: HMI Synoptic; \textit{Bottom left}: GONG Zero-point corrected; \textit{Bottom right}: GONG ADAPT Realization \#10. The green vertical line indicates the Carrington longitude of the central meridian as seen from Earth. The dotted circle marks the approximate location of the CME source region. The gray curve marks the location of the heliospheric current sheet at 2.5~R$_\odot$. All maps are saturated to ${\pm}100$~G, with red (blue) indicating positive (negative) polarity.}
 \label{fig:magnetograms_event2}
 \end{figure*}
 
 \begin{figure*}[t!]
 \resizebox{1\hsize}{!}{\includegraphics[angle=0]{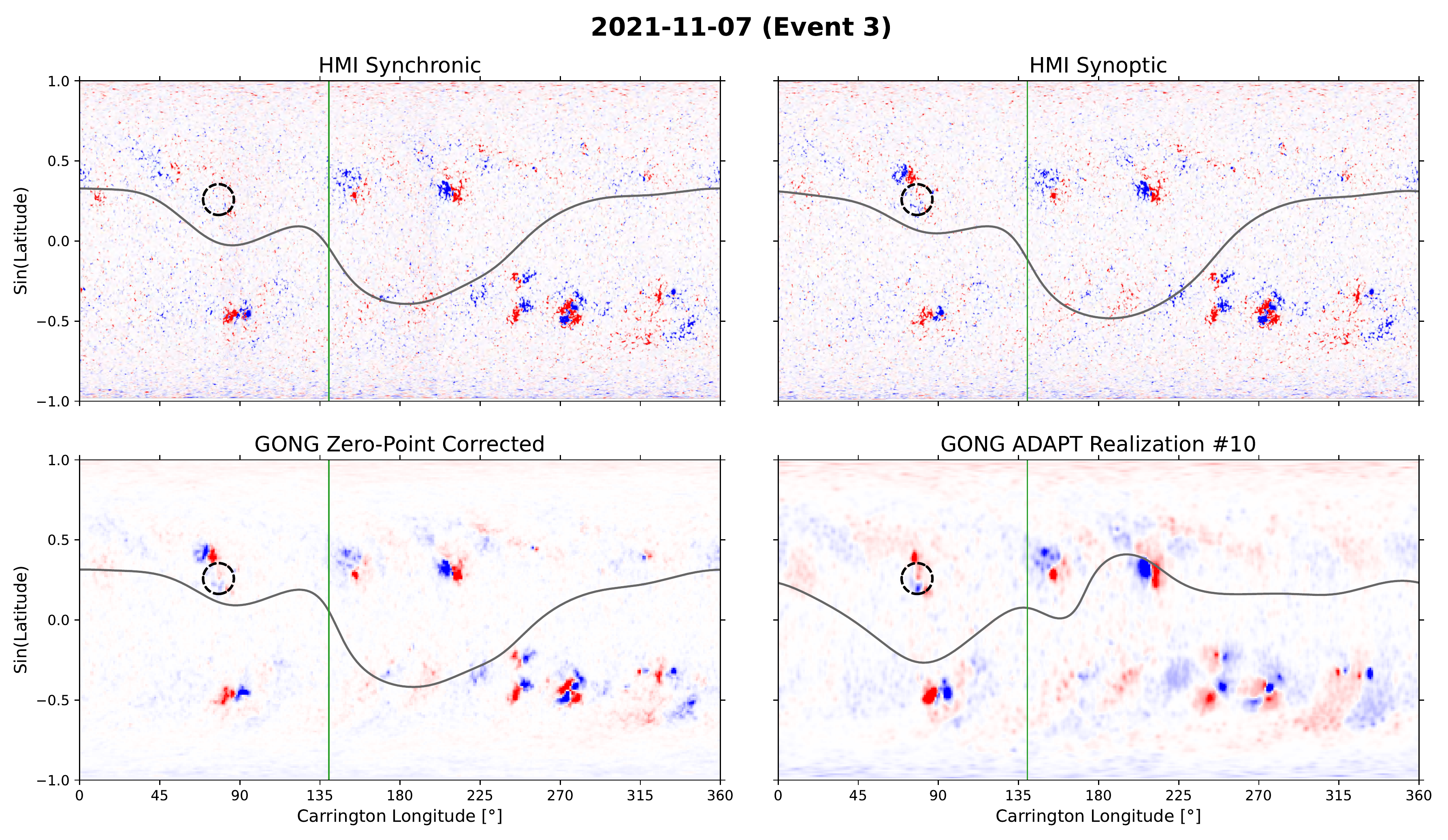}}
  \centering 
  \caption{The four magnetograms that we use as input photospheric conditions for Event~3.  \textit{Top left}: HMI Synchronic; \textit{Top right}: HMI Synoptic; \textit{Bottom left}: GONG Zero-point corrected; \textit{Bottom right}: GONG ADAPT Realization \#10. The green vertical line indicates the Carrington longitude of the central meridian as seen from Earth. The dotted circle marks the approximate location of the CME source region. The gray curve marks the location of the heliospheric current sheet at 2.5~R$_\odot$. All maps are saturated to ${\pm}100$~G, with red (blue) indicating positive (negative) polarity.}
  \label{fig:magnetograms_event3}
  \end{figure*}
  
 \begin{figure*}[t!]
 \resizebox{1\hsize}{!}{\includegraphics[angle=0]{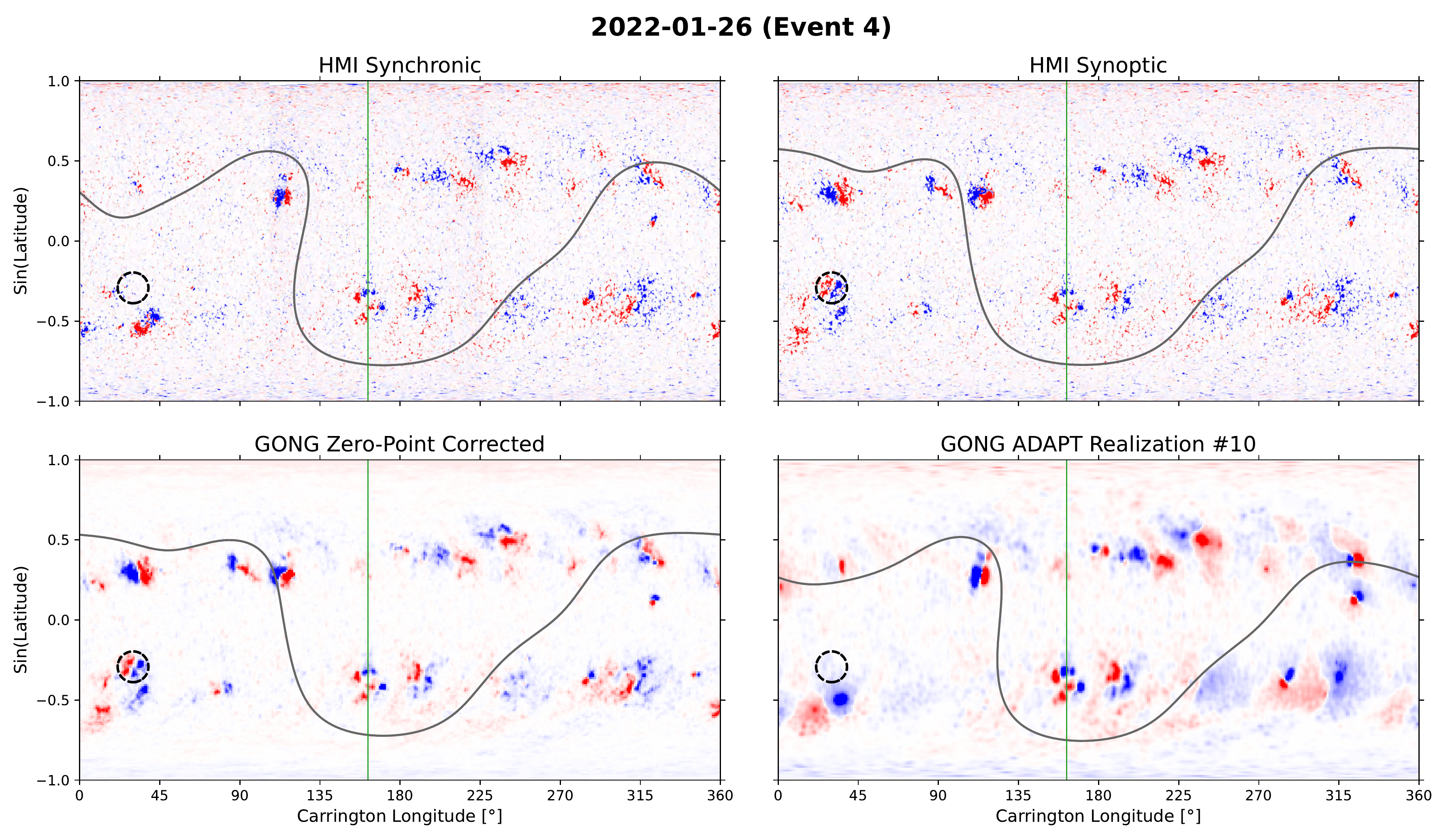}}
  \centering 
  \caption{The four magnetograms that we use as input photospheric conditions for Event~4.  \textit{Top left}: HMI Synchronic; \textit{Top right}: HMI Synoptic; \textit{Bottom left}: GONG Zero-point corrected; \textit{Bottom right}: GONG ADAPT Realization \#10. The green vertical line indicates the Carrington longitude of the central meridian as seen from Earth. The dotted circle marks the approximate location of the CME source region. The gray curve marks the location of the heliospheric current sheet at 2.5~R$_\odot$. All maps are saturated to ${\pm}100$~G, with red (blue) indicating positive (negative) polarity.}
  \label{fig:magnetograms_event4}
  \end{figure*}

In this section, we analyze the input magnetograms for each event and comment on their differences (see Sect.~\ref{subsec:magnetograms} for a description of the different datasets). Figs.~\ref{fig:magnetograms_event1}, \ref{fig:magnetograms_event2}, \ref{fig:magnetograms_event3}, and \ref{fig:magnetograms_event4} show the four magnetograms that associated with Events 1 to 4, respectively.  Across the four events, we notice that two pairs of magnetograms feature magnetic features in their images that are more consistent with each other: HMI Synchronic and GONG Adapt Realization \#10 is one pair, and HMI Synoptic and GONG Zero-point corrected is another pair.  This makes sense as these are pairs of synchronic and synoptic magnetograms, respectively.  The two GONG magnetograms also have a lower resolution than the HMI magnetograms, with magnetic elements appearing larger and less defined.    In the HMI synchronic magnetograms, it is possible to notice the two lines that mark where ``stitching'' occurs between the synchronic magnetogram snapshots and the synoptic data. The GONG Adapt Realization does not have a similar stitching line, since the photospheric field that does not face Earth is modeled via flux transport algorithms.

We also note differences in synchronic vs. synoptic magnetograms when finding active regions at the locations of the CME source regions.  In Figs.~\ref{fig:magnetograms_event1}--\ref{fig:magnetograms_event4} we include a dotted circle as an indicator of the location from where each CME launched. For Event~1, we use the Carrington coordinates provided in \citet{palmerio2021b}, since the CME was stealthy and was not associated with low-coronal signatures. For Events~2--4, we use remote-sensing extreme-ultraviolet (EUV) measurements from SDO and/or STEREO-A to locate the source active regions.

In Events 2 and 4, the identified source region matches the location of an active region only in the synoptic magnetograms.  Both synchronic magnetograms do not show a source magnetic region in the vicinity of the CME's real source as determined from EUV images.  In synchronic magnetograms, the area from the left-hand border of the magnetogram (i.e., Carrington longitude of 0$^{\circ}$) up to the ``stitching'' line includes magnetic field data from the previous Carrington rotation.  In other words, synchronic magnetograms cannot use ``future'' data, since synoptic maps are made available once per Carrington rotation, while synchronic magnetograms are updated daily. This affects magnetic elements that may be on the ``left side'' of the stitching area, in particular new flux emergence.  Notably, the ``stitching'' area does not cover the entire Earth-facing solar disk, but only ${\pm}60^{\circ}$ from the central meridian, hence active regions close to the limb of the Earth-facing solar disk may be misrepresented in synchronic magnetic maps.
  

\subsection{OSPREI Input Parameters} \label{subsec:inputs}

Once the events to be investigated in this study and their source regions have been identified, the next step is to set up the OSPREI input parameters that describe each CME as well as the background solar wind through which they propagate. Our aim is to use the same set of input parameters for each set of 64 ensemble model runs (combination of four magnetograms and 16 PFSS source surface heights, see Sect.~\ref{sec:setup}) and evaluate whether there are any significant differences and/or emerging trends. While some parameters (such as source region coordinates) are taken directly from remote-sensing and in-situ observations, some others (such as solar wind drag coefficient) are inherently less constrained by data. Since our goal in this work is not to perform forecasts, but rather to explore the photospheric and coronal input parameter space, we attempt to fine-tune these less-constrained parameters in order to match the CME arrival time at PSP. When performing the fine-tuning, we use as reference what we call the ``baseline'' run, i.e.\ the run that uses HMI synchonic magnetograms and a PFSS source surface height of 2.5\,R$_{\odot}$---which is considered to be the ``standard'' setup for OSPREI applications. Table~\ref{tab:table1}, similar to Table~1 in \citet{kay2022a}, shows the free parameters for all four events that we adjust before running OSPREI. \editone{The set of input parameters can be divided into CME and ambient medium ones, and are illustrated in more detail below together with a description of how each quantity is selected---using a combination of real measurements and best-guess approximations.}

\begin{sidewaystable*}[p]
\centering
\begin{tabular}{lccccc}
\toprule
OSPREI Input & Event 1 & Event 2 & Event 3 & Event 4 & Source\\
\midrule
\rule{0pt}{1.5ex}\\
CME parameters\\
\cline{1-1}
\noalign{\vskip 2mm}
Date & 21 June 2020 & 9 June 2021 & 7 November 2021 & 26 January 2022\\
Start time ($\mathrm{t_{0}}$) & 18:00 UT & 12:00 UT & 13:00 UT & 21:00 UT & EUV\\
Initial Carrington latitude ($\theta$) & $-18^{\circ}$ & 26$^{\circ}$ & 15$^{\circ}$ & $-17^{\circ}$ & Magnetogram\\
Initial Carrington longitude ($\phi$) & 340$^{\circ}$ & 77$^{\circ}$ & 78$^{\circ}$ & 30$^{\circ}$ & Magnetogram\\
Initial tilt ($\psi$) & 15$^{\circ}$ & 140$^{\circ}$ & 110$^{\circ}$ & 60$^{\circ}$ & Magnetogram\\
Flux rope handedness (H) & + & -- & -- & + & EUV/HHR\\
Initial nose height ($\mathrm{R_{0}}$) & 1.5\,R$_{\odot}$ & 1.1\,R$_{\odot}$ & 1.1\,R$_{\odot}$ & 1.1\,R$_{\odot}$ & EUV\\
Initial slow-rise velocity ($\mathrm{V_{R0}}$) & 30~km~s$^{-1}$ & 100~km~s$^{-1}$ & 30~km~s$^{-1}$ & 50~km~s$^{-1}$ & Estimated \\ 
Start of rapid acceleration ($\mathrm{R_{1}}$) & 4 R$_{\odot}$ & 1.3 R$_{\odot}$ & 4.8 R$_{\odot}$ & 1.1 R$_{\odot}$ & Estimated \\ 
Height of max coronal velocity ($\mathrm{R_{2}}$) & 20 R$_{\odot}$ & 5 R$_{\odot}$ & 8 R$_{\odot}$ & 3 R$_{\odot}$ & Estimated \\ 
Max coronal velocity ($\mathrm{V_{R2}}$) & 350~$\mathrm{km~s^{-1}}$ & 600~$\mathrm{km~s^{-1}}$ & 650~$\mathrm{km~s^{-1}}$ & 500~$\mathrm{km~s^{-1}}$ & GCS\\
Angular width (AW) & 30$^{\circ}$ & 35$^{\circ}$ & 30$^{\circ}$ & 45$^{\circ}$ & GCS\\
Perpendicular AW ($\mathrm{AW_{\perp}}$) & 10$^{\circ}$ & 15$^{\circ}$ & 15$^{\circ}$ & 20$^{\circ}$ & GCS\\
Flux rope initial magnetic field (B$_{0}$) & 1200~nT & 4000~nT & 3200~nT & 2500~nT & Estimated\\
Flux rope initial temperature (T$_{0}$) & $2.11\times10^{6}$~K & $4.44\times10^{6}$~K & $2.75\times10^{6}$~K & $2.31\times10^{6}$~K & Estimated\\
CME mass ($\mathrm{M_{CME}}$) & 2 $\mathrm{\times~10^{15}~g}$ & 1 $\mathrm{\times~10^{16}~g}$ & 7 $\mathrm{\times~10^{15}~g}$ & 6 $\mathrm{\times~10^{15}~g}$ & Estimated\\
Coronal axis aspect ratio ($\mathrm{\delta_{Ax}}$) & 0.6 & 0.6 & 0.6 & 0.7 & Fine-tuned\\
Coronal cross-section aspect ratio ($\mathrm{\delta_{CS}}$) & 1 & 1 & 1 & 1 & Default\\
Adiabatic index ($\mathrm{\gamma}$) & 1.66 & 1.33 & 1.5 & 1.66 & Fine-tuned\\ 
Interplanetary expansion factor ($\mathrm{f_{exp}}$) & 0.50 & 0.25 & 0.50 & 0.85 & Fine-tuned\\ 
\rule{0pt}{1.5ex}\\
Ambient medium parameters\\
\cline{1-1}
\noalign{\vskip 2mm}
Drag coefficient (C$_\mathrm{D}$) & 0.9 & 0.8 & 1 & 1 & Fine-tuned\\ 
Solar wind speed (V$_\mathrm{SW}$) & 300~$\mathrm{km~s^{-1}}$ & 350~$\mathrm{km~s^{-1}}$ & 300~$\mathrm{km~s^{-1}}$ & 350~$\mathrm{km~s^{-1}}$ & PSP data\\
Solar wind density (N$_\mathrm{SW}$) & 25~$\mathrm{cm^{-3}}$ & 25~$\mathrm{cm^{-3}}$ & 55~$\mathrm{cm^{-3}}$ & 20~$\mathrm{cm^{-3}}$ & PSP data\\
Solar wind magnetic field (B$_\mathrm{SW}$) & 10~nT & 5~nT & 20~nT & 5~nT & PSP data\\
Solar wind temperature (T$_\mathrm{SW}$) & $2.5\times10^{4}$~K & $3.0\times10^{4}$~K & $7.0\times10^{4}$~K & $5.0\times10^{4}$~K & PSP data\\
Initial s/c latitude (Carrington) & 2.0$^{\circ}$ & 3.4$^{\circ}$ & 3.8$^{\circ}$ & 3.8$^{\circ}$ & PSP ephemeris\\
Initial s/c longitude (Carrington) & 327.1$^{\circ}$ & 74.0$^{\circ}$ & 75.0$^{\circ}$ & 12.6$^{\circ}$ & PSP ephemeris\\
Initial s/c heliocentric distance & 96.8 R$_{\odot}$ & 163.9 R$_{\odot}$ & 96.6 R$_{\odot}$ & 147.3 R$_{\odot}$ & PSP ephemeris\\
\bottomrule
\end{tabular}
\caption{Full list of OSPREI free Parameters for the four CME events. For a more robust explanation of each parameter, please see \citet{kay2022a}. Latitudes and longitudes are expressed in Carrington coordinates. The ``Source'' column indicates where the value/information was sourced from.}
\normalsize \label{tab:table1} 
\end{sidewaystable*}

From EUV \editone{observations, we identify the source region and establish} the CME start time \editone{($\mathrm{t_{0}}$), by evaluating the first signs of any associated motion}. \editone{Using the identified source and} magnetogram data, we determine the \editone{CME} initial latitude and longitude \editone{($\theta$ and $\phi$ respectively, i.e., the Carrington coordinates of the corresponding source region)}, as well as the tilt \editone{($\psi$)}. The tilt is the angle that describes the direction of the flux rope axis and is defined to move counterclockwise from the solar west direction. \editone{We initially determine the tilt from the orientation of the local polarity inversion line \citep[e.g.,][]{marubashi2015}, and then we resolve the $180^{\circ}$ ambiguity by means of the flux rope handedness (H, also known as chirality), assuming that the MFR axial field points from the positive magnetic field polarity to the negative one.} The flux rope handedness of each CME is determined from EUV observations \citep[e.g.,][]{palmerio2017}, where possible (Event~3), and with the hemispheric helicity rule \citep[\editone{HHR;}][]{pevtsov2014}, also known as the Bothmer--Schwenn scheme \citep[from][]{bothmer1998}, otherwise (Events~1, 2, and 4). \editone{According to the HHR, which is of statistical nature, CMEs from the northern (southern) hemisphere tend to be characterized by a left- (right-) handed chirality. We note that for Event~3, the handedness determined from remote-sensing observations is in agreement with the HHR.}

\editone{Another OSPREI parameter to define is the height of the CME nose at the start of the run, $\mathrm{R_{0}}$.} Event~1 is a streamer-blowout stealth CME that was observed in STEREO/EUVI-A \editone{off-limb} images to initiate high up in the corona, \editone{at around 1.5~R$_{\odot}$}. Events~2, 3, and 4 \editone{were all active-region eruptions with clear on-disk signatures, indicating that they originated much lower in the corona. For these events, we use the OSPREI default value of 1.1~R$_{\odot}$, which allows for inclusion of low-coronal effects but is not so low as to be sensitive to ringing effects that the PFSS model occasionally suffers near high-intensity active regions. From the height of $\mathrm{R_{0}}$, CMEs in OSPREI initially move at their slow-rise velocity ($\mathrm{V_{R0}}$) until their reach a point ($\mathrm{R_{1}}$) from which they start their acceleration phase. During the acceleration phase, the CME speed increases linearly from $\mathrm{V_{R0}}$ up to $\mathrm{V_{R2}}$, i.e.\ the maximum coronal velocity, which is attained at a height $\mathrm{R_{2}}$. After $\mathrm{R_{2}}$, it is assumed that CMEs continue their coronal and interplanetary propagation at a constant speed until the outer boundary of the coronal domain (set in this work at 20~R$_{\odot}$), when OSPREI begins to simulate drag and other interplanetary processes that affect the CME speed. These steps are meant to emulate the ``classic'' kinematic evolution of CMEs, consisting of the an initiation phase, an impulsive acceleration phase, and a propagation phase \citep[e.g.,][]{zhang2001}. We select values for $\mathrm{V_{R0}}$, $\mathrm{R_{1}}$, and $\mathrm{R_{2}}$ via rough estimates of the CME kinematics as observed in (off-limb) EUV and coronagraph imagery.} 

\editone{The CME maximum coronal speed ($\mathrm{V_{R2}}$), on the other hand, together with both angular-width parameters (AW and AW$_{\perp}$), are determined using the Graduated Cylindrical Shell \citep[GCS;][]{thernisien2011} model applied to nearly-simultaneous coronagraph images from SOHO and STEREO-A. The GCS geometry is meant to represent the flux rope morphology of CMEs, resulting in a toroidal body and two legs curving toward the Sun, reminiscent of a croissant. It follows that this structure has an elliptical projection onto the plane perpendicular to the radial direction at the CME nose, and the angular-width parameters are represented by the semi-major (AW) and semi-minor (AW$_{\perp}$) axes of such an ellipse, corresponding to the CME face-on and edge-on angular widths, respectively. The maximum coronal speed $\mathrm{V_{R2}}$ is simply derived from the positions of the CME apex given by two GCS reconstructions separated by 1~hour.} 

\editone{The CME initial magnetic field (B$_{0}$) and temperature (T$_{0}$) are loosely estimated based on previous OSPREI work, under the assumption that these input quantities tend to vary with the overall scale of the CME (e.g., size and speed), with more extreme events being generally associated with higher values. The CME mass ($\mathrm{M_{CME}}$) is estimated ``by eye'' based on the brightness of each event in coronagraph imagery, using a lower-bound value of $2\times10^{15}$~g for Event~1, i.e.\ the faintest in our set \citep[using the estimate of][]{palmerio2021b} and an upper-bound value of $10^{16}$~g for Event~2, i.e.\ the brightest in our set. In fact, \citet{Gopalswamy2005} found in a study of all CMEs observed over almost a full month that the distribution of masses tends to display a double peak at ${\sim}10^{15}$~g and ${\sim}10^{16}$~g, with extremely narrow (${\sim}10^{\circ}$ width) CMEs featuring values of the order of $10^{14}$~g and extreme space weather events (such as those associated with the 2003 Halloween storm) surpassing $10^{17}$~g. We note that precise determination of the last three input parameters (B$_{0}$, T$_{0}$, and $\mathrm{M_{CME}}$) in the context of space weather forecasting is a critical subject of ongoing studies \citep[e.g.,][]{vourlidas2019}. For example, existing works have attempted to estimate the CME magnetic field strength from radio measurements \citep[e.g.,][]{carley2017}, or the CME mass from white-light coronagraph images \citep[e.g.,][]{colaninno2009, temmer2021} and EUV dimmings \citep[e.g.,][]{lopez2019}.}

\editone{The geometry of the MFR embedded in the modeled CMEs is described by the coronal axis aspect ratio ($\mathrm{\delta_{Ax}}$) and cross-section aspect ratio ($\mathrm{\delta_{CS}}$). As mentioned in Sect.~\ref{subsec:osprei}, OSPREI uses the EC analytical flux rope model of \citet{nieves-chinchilla2018}, and the implemented MFR geometry consists of a toroidal axis defined by an ellipse with semi-major and semi-minor lengths L$_{\perp}$ and L$_\mathrm{r}$, and an elliptical cross-section with semi-major and semi-minor lengths r$_{\perp}$ and r$_\mathrm{r}$ \citep[more information can be found in][]{kay2021a}. The ratios mentioned above correspond to $\mathrm{\delta_{Ax}}$ = L$_\mathrm{r}$/L$_{\perp}$ and $\mathrm{\delta_{CS}}$ = r$_\mathrm{r}$/r$_{\perp}$. In this work, we keep all inputs for $\mathrm{\delta_{CS}}$ to their OSPREI default value of 1, corresponding to a circular cross-section of the MFR, while for $\mathrm{\delta_{Ax}}$ we use values in the range 0.6--0.7, which emulate the ``classic'' geometry of flux ropes that assumes an axis that is elongated in one direction \citep[e.g.,][]{krall2006}.}

\editone{The last two CME parameters, i.e.\ the adiabatic index ($\mathrm{\gamma}$) and the interplanetary expansion factor ($\mathrm{f_{exp}}$), control the CME propagation due to internal forces. The adiabatic index allows for the CME thermal expansion to vary between isothermal ($\mathrm{\gamma}$ = 1) and adiabatic ($\mathrm{\gamma}$ = 1.67), while the expansion factor describes the CME initial velocity decomposition, or how the propagation speed translates into expansion speed. For $\mathrm{f_{exp}}$ = 0, the CME experiences fully self-similar behavior; for $\mathrm{f_{exp}}$ = 1, the CME undergoes fully convective expansion. $\mathrm{f_{exp}}$ only sets the expansion speeds at the beginning of the interplanetary portion, beyond this first step the expansion evolves according to the interplanetary (drag, magnetic, and thermal) forces. We set the values for $\mathrm{\gamma}$ and $\mathrm{f_{exp}}$ to explore different propagation scenarios across the four events under study, with minor adjustments to ultimately yield better CME arrival times for the baseline run.}

\editone{Finally, we describe the ambient medium parameters used in OSPREI. The (dimensionless) drag coefficient ($\mathrm{C_{D}}$) quantifies the external drag exerted on the CME by the background solar wind \citep{cargill2004}. We initially use the default OSPREI value for $\mathrm{C_{D}}$ of 1, and we slightly vary it in order to obtain a better CME arrival time for the baseline run, where necessary. This results in $\mathrm{C_{D}}$ values of 0.9 and 0.8 for Events~1 and 2, respectively.} We find the ambient velocity ($\mathrm{V_{SW}}$), density ($\mathrm{N_{SW}}$), magnetic field ($\mathrm{B_{SW}}$), and temperature ($\mathrm{T_{SW}}$) by examining the PSP solar wind data for $\sim$1 day before the CME impact and by taking an average value for each parameter. The only exception is the background speed for Event~3, which had no available data, and for which we \editone{use} a slow solar wind speed of 300~km~s$^{-1}$ \editone{based on the results of \citet{mcgregor2011}, who found that the slow wind close to solar minimum peaks around this value at 0.3--0.4~au (PSP was at ${\sim}$0.45~au during Event~3) and is generally slower than the ambient speed at 1~au (which peaks at 350~km~s$^{-1}$)}. Table~\ref{tab:table1} also shows the initial PSP latitude, longitude, and heliocentric distance, i.e.\ the position of PSP at the CME eruption time. In addition to these values, we load onto OSPREI the PSP ephemeris throughout the simulated period, meaning that the synthetic trajectory through each CME encounter corresponds to the actual time-dependent PSP orbit.


\section{Results} \label{sec:results}

After the ensemble OSPREI runs have been performed for each event, we compare ForeCAT (coronal deflections and rotations until 20\,R$_{\odot}$) and FIDO (synthetic in-situ profiles along the PSP trajectory) results to visualize the model's sensitivity to the different input magnetograms and PFSS source surface heights. We compare the FIDO profiles with PSP data, specifically magnetic field measurements from the fluxgate magnetometer part of the FIELDS \citep{bale2016} investigation and plasma measurements from the Solar Probe Cup \citep[SPC;][]{case2020} part of the Solar Wind Electrons Alphas and Protons \citep[SWEAP;][]{kasper2016} suite. Summary plots for each CME are shown in Figs.~\ref{fig:summaryplot1}, \ref{fig:summaryplot2}, \ref{fig:summaryplot3}, and \ref{fig:summaryplot4} for Events~1, 2, 3, and 4, respectively. In these figures, each input magnetogram is represented by one color, with the thicker line showcasing the corresponding $\mathrm{R_{ss}}=2.5$\,R$_{\odot}$ case and the thinner lines representing results at all other PFSS source surface heights.

\subsection{Event 1}

\begin{figure*}[p]
\centering 
 \includegraphics[scale=0.27]{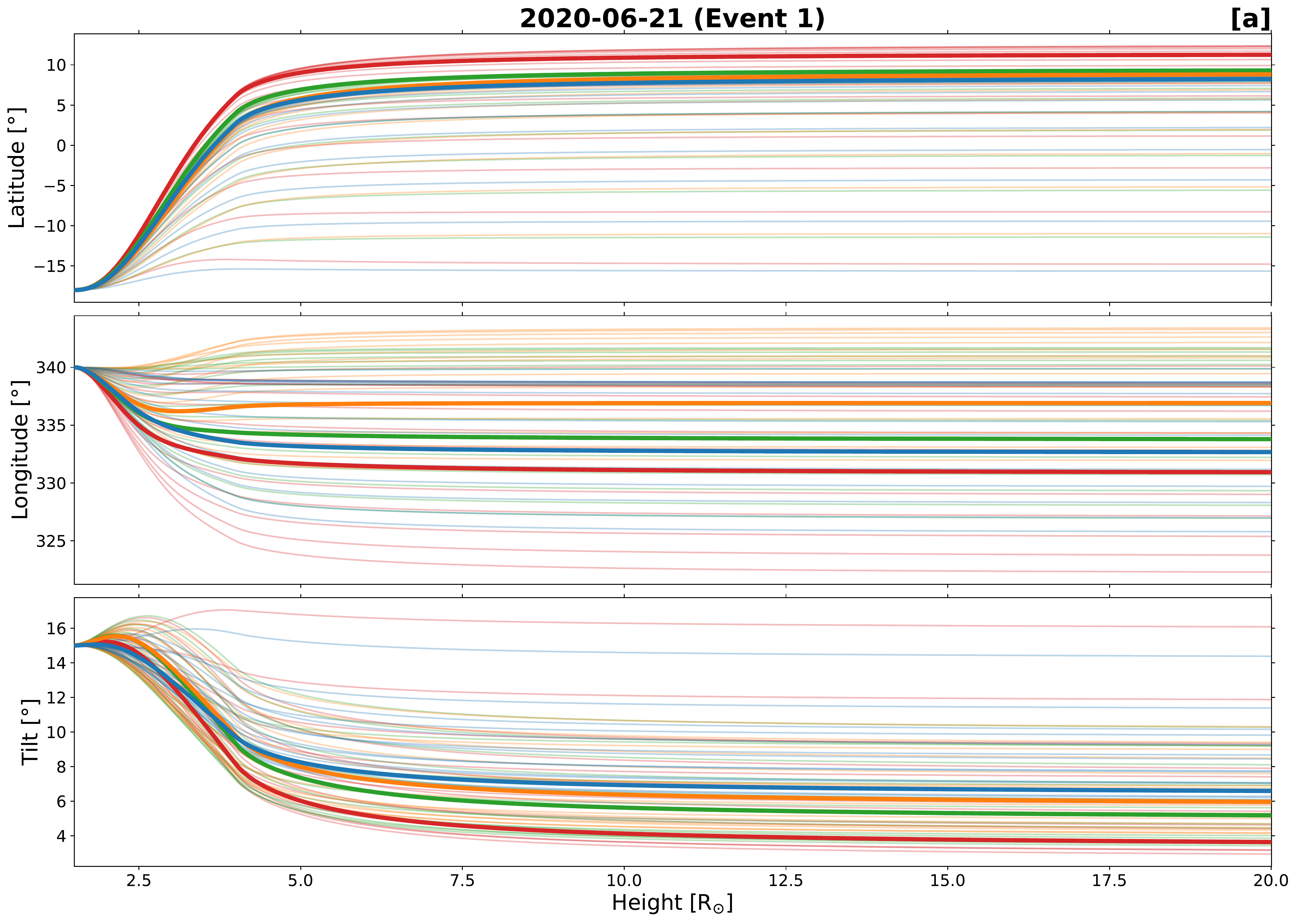}\\
 \vspace*{.1in}
 \includegraphics[scale=0.27]{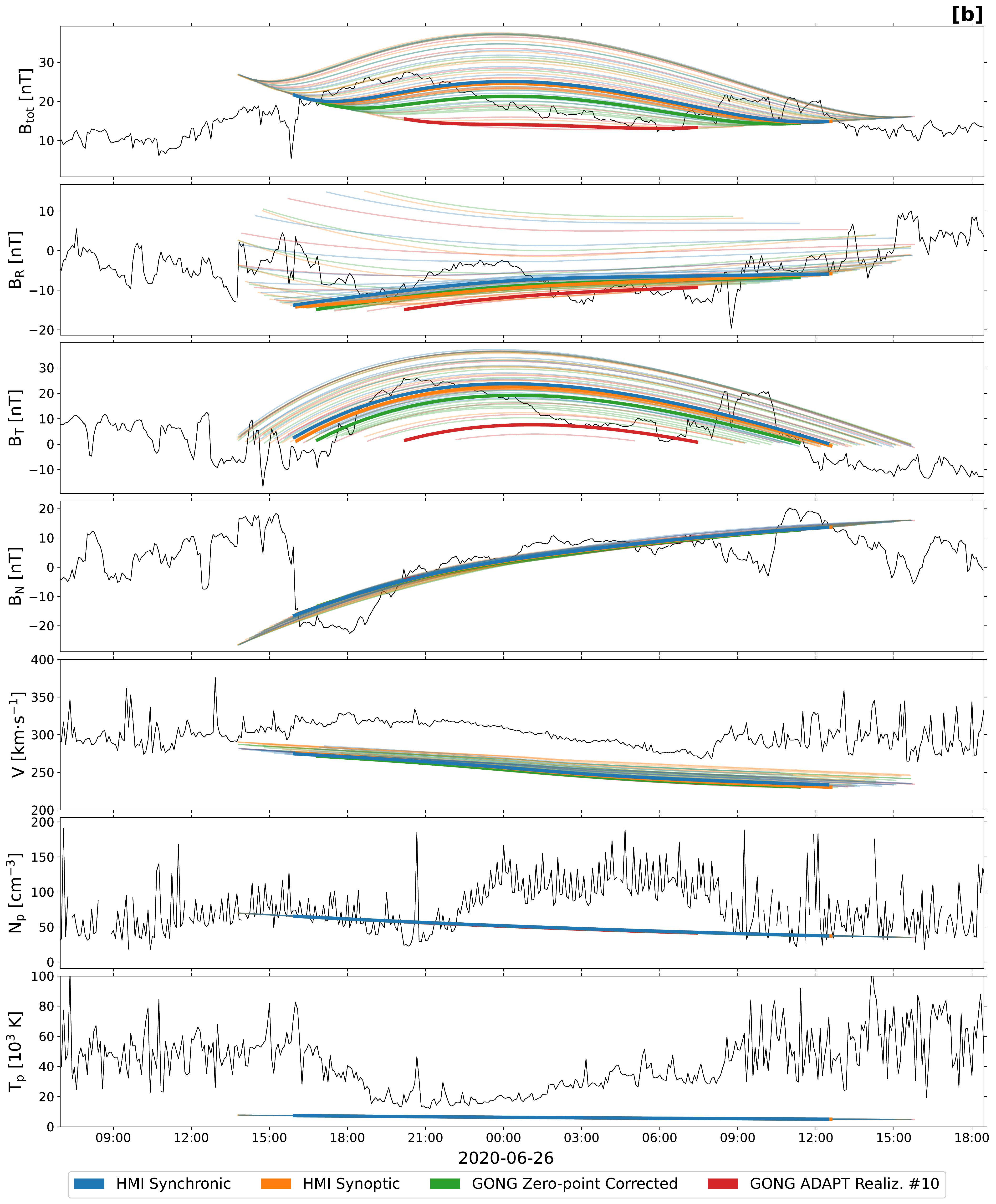}
  \captionsetup{justification=centering}
  \caption{ForeCAT and FIDO results for Event 1 plotted in panels (a) and (b), respectively.  The FIDO results are shown against PSP magnetic field and plasma data. The thicker contours represent the model output using a PFSS with $\mathrm{R_{ss}} = 2.5$~R$_{\odot}$.  The thinner contours represent all other PFSS source surface heights.  The four different colors correspond to different input magnetograms.  If a given combination of input magnetogram and PFSS source surface height resulted in a ``miss'' according to OSPREI, it was not included in the FIDO plot (see text for details).}
  \label{fig:summaryplot1}
  \end{figure*}

In Event~1 (front-sided CME as seen from Earth, Fig.~\ref{fig:summaryplot1}), there are relatively small differences between the four magnetograms for fixed $\mathrm{R_{ss}}=2.5$\,R$_{\odot}$ in the ForeCAT results. Different PFSS source surface heights result in a wider spread in both CME deflections and rotations. Latitude and longitude are more severely affected by the PFSS source surface height: The latitudes at 20\,R$_{\odot}$ span a range of ${\sim}25^{\circ}$, while the longitudes span a range of ${\sim}20^{\circ}$. The final tilts spread over a smaller range (${\sim}12^{\circ}$) that is due to a few outliers, since the majority of them are clustered around the corresponding $\mathrm{R_{ss}}=2.5$\,R$_{\odot}$ profiles.

Examining FIDO results, we note that all combinations of magnetograms and PFSS source surface heights result in a more or less accurate representation of the CME MFR, especially in the B$\mathrm{_{T}}$ (rotating from negative to positive) and B$\mathrm{_{N}}$ (positive throughout the rope) components. The B$\mathrm{_{R}}$ profiles feature larger differences, with approximately half of the runs capturing its negative trend. This is most likely a direct result of the different final latitudes in ForeCAT: The runs that ended with a positive (negative) latitude in the corona later crossed the CME below (above) its central axis in situ (we remark that all runs featured a tilt at 20\,R$_{\odot}$ in the range 4--$16^{\circ}$, indicating a low-inclination MFR with its axis nearly parallel to the solar equator). Most notably, several runs do not result in an impact at PSP---specifically, all the runs with $\mathrm{R_{ss}} = 1.5$\,R$_{\odot}$, the HMI Synchronic and GONG ADAPT runs with $\mathrm{R_{ss}} = 1.6$\,R$_{\odot}$, and the GONG ADAPT runs with $\mathrm{R_{ss}}$ in the range 2.7--3.0\,R$_{\odot}$. Of these, the runs characterized by the smaller source surface heights were those featuring the smallest latitudinal deflections in ForeCAT, suggesting that the CME remained too south of PSP to produce an impact. The GONG ADAPT runs with larger source surface heights, on the other hand, were the ones featuring the most prominent longitudinal deflections in the solar corona, up to ${>}15^{\circ}$ away from the CME source region (we remark that the CME was set to have a half-angular width of $30^{\circ}$ along its major axis, see Table~\ref{tab:table1}). Finally, we do not note significant differences in the modeled plasma parameters, i.e.\ solar wind speed, density, and temperature.

\subsection{Event 2}

\begin{figure*}[p]
\centering 
 \includegraphics[scale=0.27]{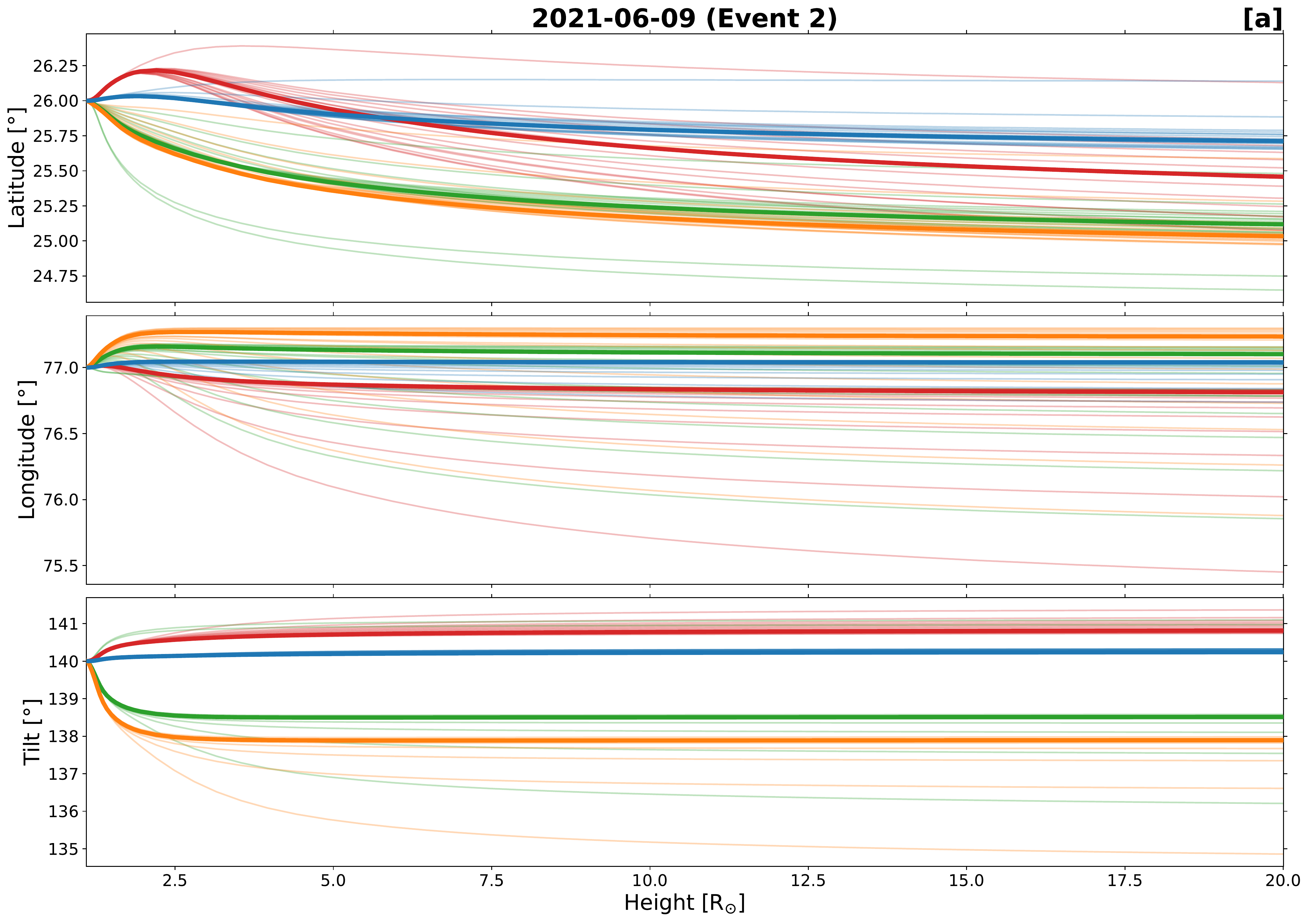}\\
 \vspace*{.1in}
 \includegraphics[scale=0.27]{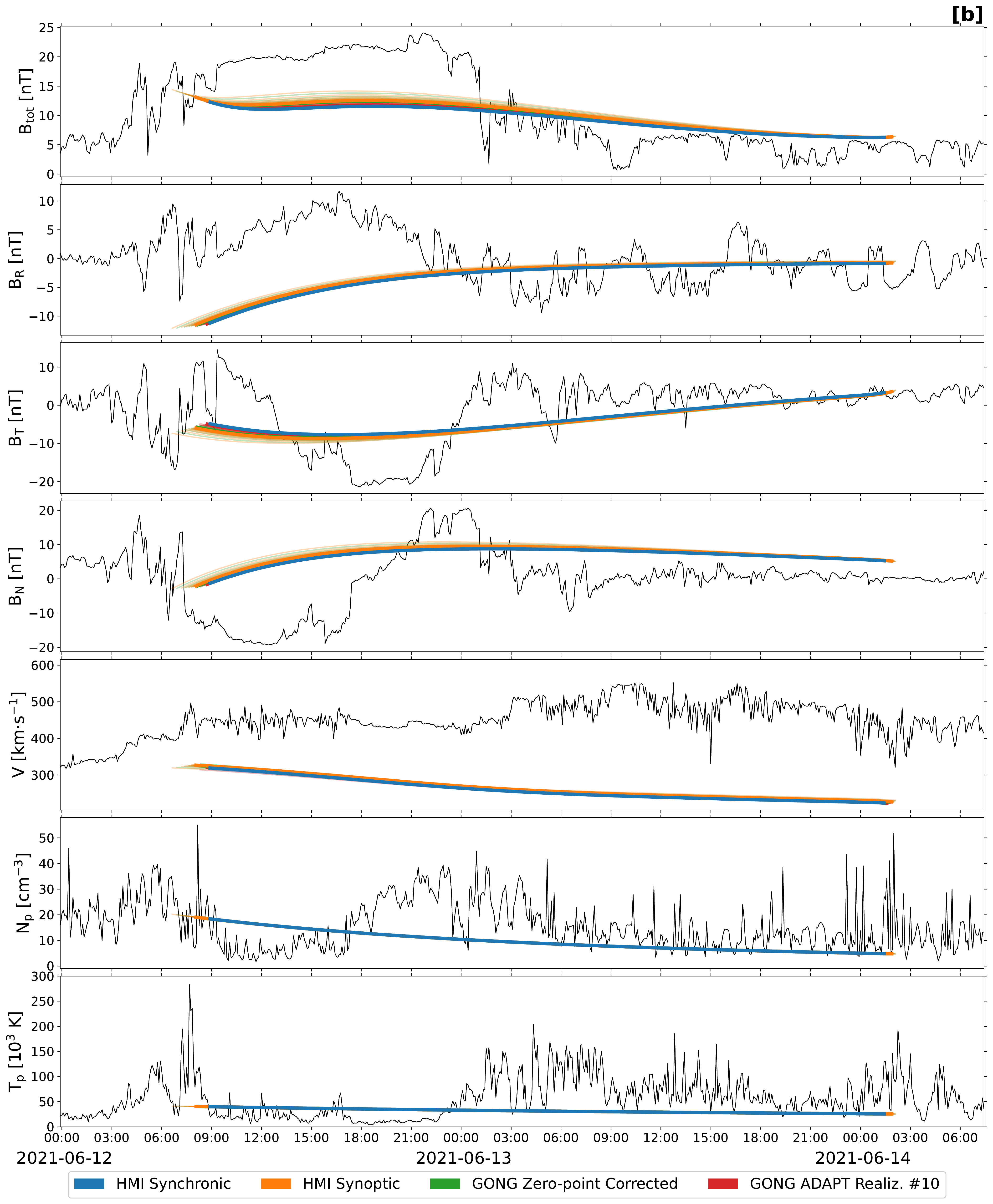}
  \captionsetup{justification=centering}
  \caption{ForeCAT and FIDO results for Event 2 plotted in panels (a) and (b), respectively. The FIDO results are shown against PSP magnetic field and plasma data. The thicker contours represent the model output using a PFSS with $\mathrm{R_{ss}} = 2.5$~R$_{\odot}$.  The thinner contours represent all other PFSS source surface heights.  The four different colors correspond to different input magnetograms.}
  \label{fig:summaryplot2}
  \end{figure*}

In Event~2 (west-limb CME as seen from Earth, Fig.~\ref{fig:summaryplot2}), there are minimal variations in the ForeCAT results at 20\,R$_{\odot}$. The CME is modeled to experience virtually no deflections regardless of the choice of input magnetogram and PFSS source surface height, and the final latitudes and longitudes spread over ${<}2^{\circ}$ across all the 64 runs. The tilt values span a range of ${\sim}6^{\circ}$---due to three outliers, since the majority of them spread over ${\sim}3^{\circ}$ only---and all indicate an MFR inclined by ${\sim}40^{\circ}$ with respect to the solar equator.

Regarding FIDO results, there are no major differences in the predicted MFR configuration, even with the varying PFSS source surface height, and all 64 ensemble runs result in an impact at PSP. Interestingly, none of the magnetic field components is modeled by OSPREI particularly well in the case of this event, and the duration of the flux rope is also overestimated by almost 24 hours. The positive B$\mathrm{_{R}}$ profile is PSP data is missed in all the runs, which predicts a predominantly-negative radial field. Results are slightly better for B$\mathrm{_{T}}$, where the negative sign is correctly captured, and for B$\mathrm{_{N}}$, in which however the duration of the negative field is considerably underestimated. Nevertheless, all the runs produce very similar results for each of the magnetic field components. We also do not note significant differences in the modeled plasma parameters, i.e.\ solar wind speed, density, and temperature. Regardless of the accuracy of the model in predicting the CME magnetic configuration, in the case of this event the MFR profiles appear to be basically unaffected by the input photospheric and coronal magnetic fields.

\subsection{Event 3}

\begin{figure*}[p]
\centering 
 \includegraphics[scale=0.27]{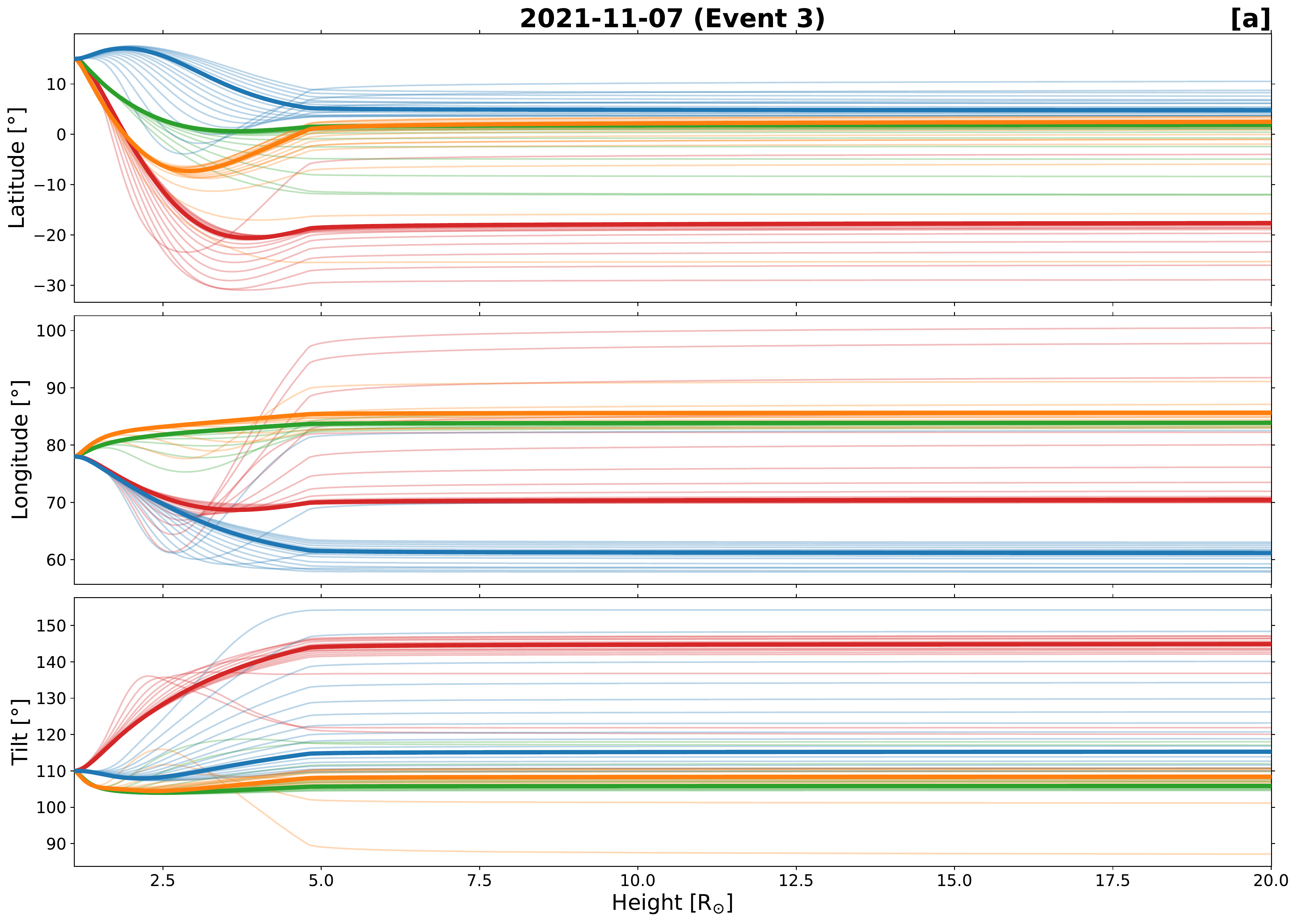}\\
 \vspace*{.1in}
 \includegraphics[scale=0.27]{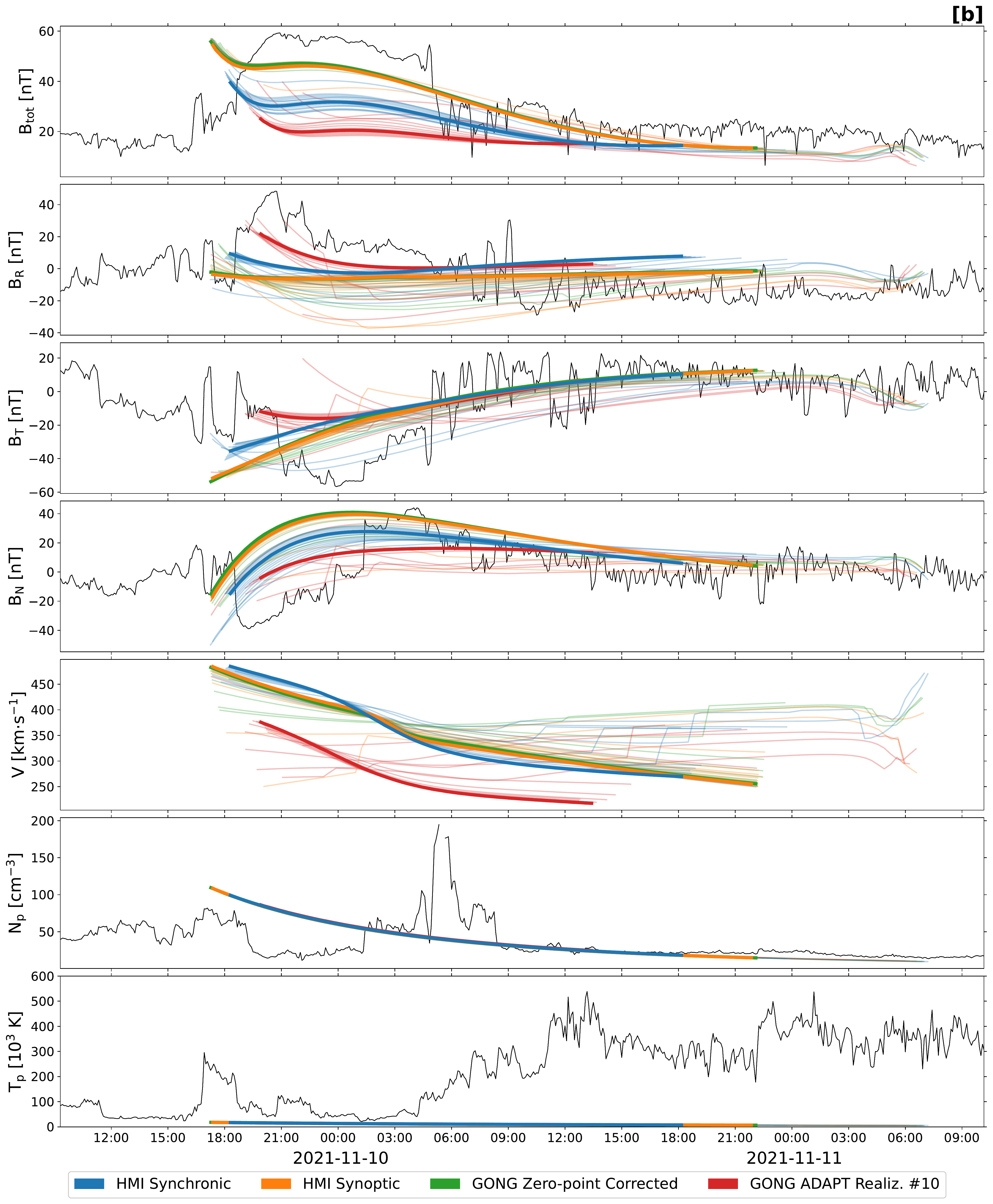}
  \captionsetup{justification=centering}
  \caption{ForeCAT and FIDO results for Event 3 plotted in panels (a) and (b), respectively. The FIDO results are shown against PSP magnetic field and plasma data.  The thicker contours represent the model output using a PFSS with $\mathrm{R_{ss}} = 2.5$~R$_{\odot}$.  The thinner contours represent all other PFSS source surface heights.  The four different colors correspond to different input magnetograms.}
  \label{fig:summaryplot3}
  \end{figure*}

In Event~3 (east-limb CME as seen from Earth, Fig.~\ref{fig:summaryplot3}), we note a high sensitivity to the input magnetogram and PFSS source surface height. In ForeCAT, in the range 1.1--5~R$_{\odot}$, the CME undergoes very different coronal deflections depending on the input magnetogram and PFSS source surface height. While the CME is seen to deflect southward in all runs, the latitudes at 20\,R$_{\odot}$ are spread over ${\sim}40^{\circ}$. Considering the $\mathrm{R_{ss}}=2.5$\,R$_{\odot}$ cases, we note that the GONG ADAPT run reaches a final latitude of approximately $-20^{\circ}$, while the three remaining magnetograms result in a ${\sim}0^{\circ}$ latitude. Significant differences are also seen in the longitudes, with all the synoptic runs (regardless of source surface height) experiencing minimal deflections, all the HMI Synchronic runs deflecting by ${\sim}20^{\circ}$ eastward, and the GONG ADAPT runs deflecting initially eastward in all cases, but then resulting in either eastward or westward deflections at 20\,R$_{\odot}$ depending on the PFSS source surface heigh. As for the latitudes, the final longitudes are overall spread over ${\sim}40^{\circ}$. Furthermore, with GONG ADAPT at $\mathrm{R_{ss}}=2.5$\,R$_{\odot}$, the CME rotates by about ${\sim}30^{\circ}$, while the other magnetograms induce minimal tilt changes---although there appears to be a large sensitivity to the PFSS source surface height in the HMI Synchronic case. The final tilts at 20\,R$_{\odot}$ are spread over ${\sim}60^{\circ}$ across all runs. Overall, ForeCAT is very sensitive to both input magnetogram and PFSS surface height for this event, with a large distribution of final latitudes, longitudes, and tilts.

Examining FIDO results (where all 64 runs produce an impact at PSP), it is clear that the differences persist, as it is possible to observe clear changes in the magnetic configuration of the modeled MFR. Starting with the total field, B$\mathrm{_{tot}}$, we note that there is a wide range of maximum field strengths, from 20~nT to almost 60~nT based on the input magnetogram and PFSS source surface height being used. B$\mathrm{_{R}}$ shows a similar spread of field strengths. The B$\mathrm{_{T}}$ profile is generally well reproduced by most runs, although the magnitude of the negative dip is better captured by the synchronic magnetograms and by a only a few of the remaining cases. The negative-to-positive trend of B$\mathrm{_{N}}$ is missed by most runs, although the profile is generally better captured in a few HMI Synchronic and GONG ADAPT cases---these are the runs that were associated with the largest counterclockwise deflections in ForeCAT. In terms of flux rope duration, all combinations tend to produce overestimates by at least six~hours, with a subset of runs extending up to ${\sim}$24~hours beyond the observed ejecta trailing edge. Unfortunately, there are no PSP solar wind speed data available for this event, but we note that the modeled speed profiles again depend on what magnetogram and PFSS source surface height were input into OSPREI.  Additionally, for some source surface heights, the velocity exhibits a ``stair-stepping'' behavior towards the end of the CME, and some combinations produce sharp velocity increases or decreases at the end of the flux rope. We do not note significant differences in the modeled solar wind density and temperature between magnetograms.

\subsection{Event 4}

\begin{figure*}[p]
\centering 
 \includegraphics[scale=0.27]{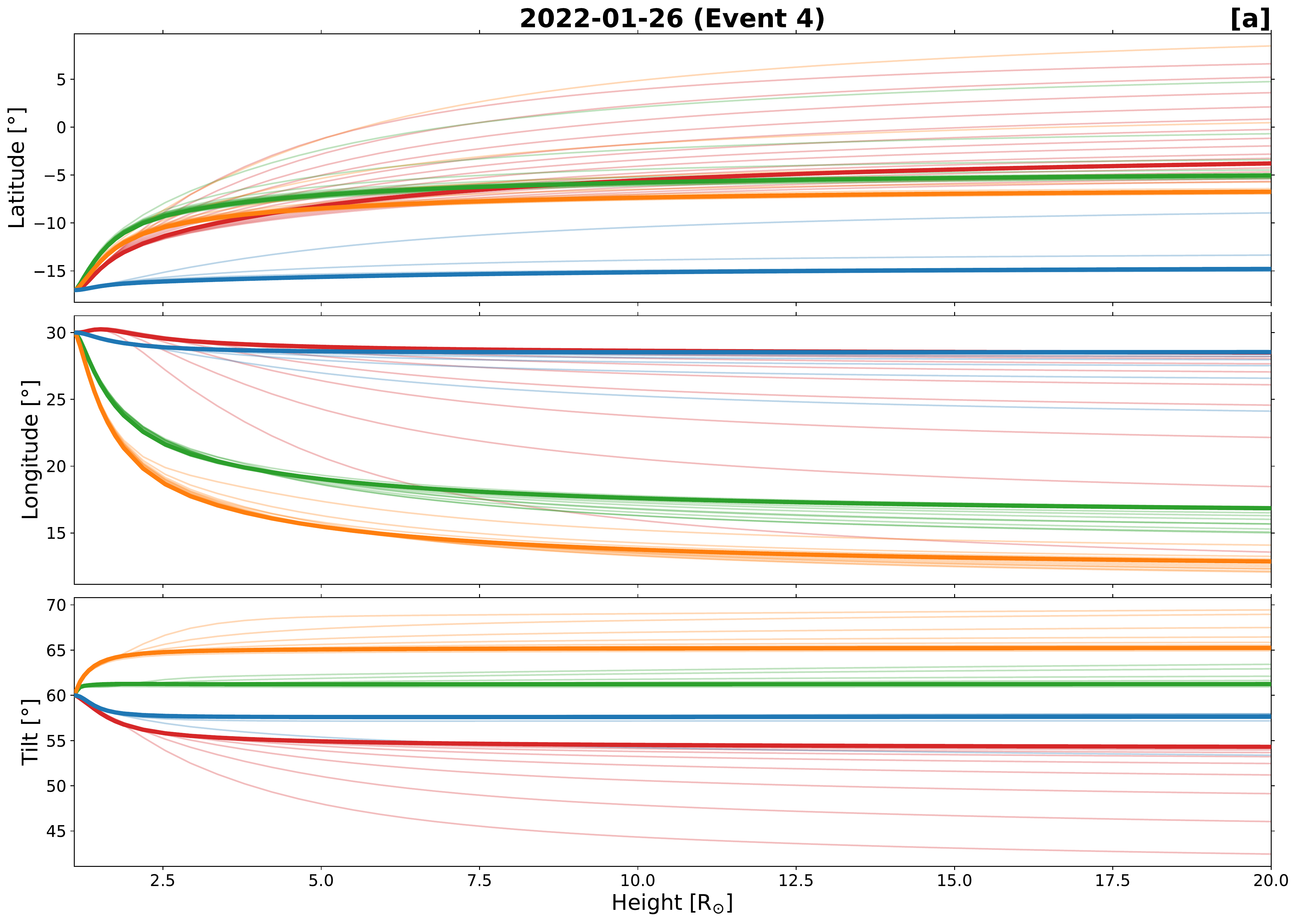}\\
 \vspace*{.1in}
 \includegraphics[scale=0.27]{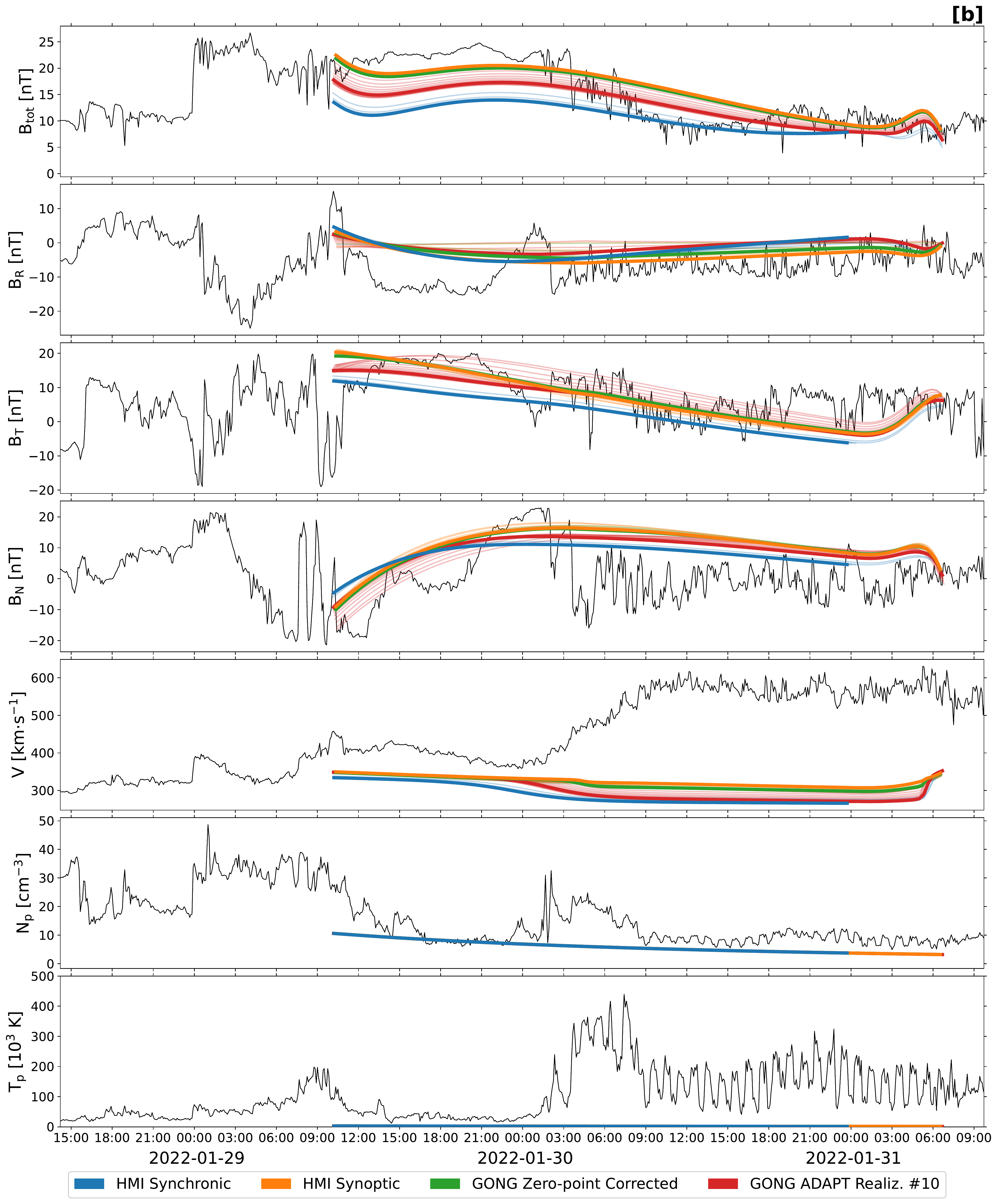}
  \captionsetup{justification=centering}
  \caption{ForeCAT and FIDO results for Event 4 plotted in panels (a) and (b), respectively.  The FIDO results are shown against PSP magnetic field and plasma data. The thicker contours represent the model output using a PFSS with $\mathrm{R_{ss}} = 2.5$~R$_{\odot}$.  The thinner contours represent all other PFSS source surface heights.  The four different colors correspond to different input magnetograms.}
  \label{fig:summaryplot4}
  \end{figure*}

In Event~4 (far-sided CME as seen from Earth, Fig.~\ref{fig:summaryplot4}), looking at ForeCAT model results, there are more or less significant differences between the input magnetograms with no particular pattern. While results at 20\,R$_{\odot}$ appear rather sensitive to the input photospheric magnetic field---with final spreads in latitude of ${\sim}25^{\circ}$, in longitude of ${\sim}15^{\circ}$, and in tilt of ${\sim}30^{\circ}$---there are generally fewer differences within a single magnetogram with respect to PFSS source surface height adjustments. The only exception is GONG ADAPT, which shows larger deviations from the ``traditional'' $\mathrm{R_{ss}}=2.5$\,R$_{\odot}$ case at other heights.

Examining FIDO (where all 64 runs produce an impact at PSP), the resulting predicted MFR structure agrees relatively well with the PSP measurements for all runs, with the largest differences observed in modeled B$\mathrm{_{tot}}$ for the HMI Synchronic runs, which underestimate the total field strength of the CME by ${\sim}10$~nT.  Most combinations include an anomalous rotation in the magnetic field components toward the end of the flux rope, although a few PFSS source surface heights paired with the HMI Synchronic magnetogram do not display this characteristic. Generally, all runs display a similar magnetic configuration of the modeled flux rope, despite the more or less large differences in the ForeCAT outputs. The MFR duration is overestimated by ${\sim}24$~hours in all runs. Finally, we do not note significant differences in the modeled plasma parameters, i.e.\ solar wind speed, density, and temperature.

\section{Discussion} \label{sec:discuss}

The results presented in Sect.~\ref{sec:results} show that there are moderate dependencies of the OSPREI model to the input photospheric and coronal conditions. OSPREI has been used and tested mostly in the case of Earth-impacting events, hence the combination of HMI Synchronic magnetogram and a PFSS source surface height of 2.5~R$_{\odot}$ (referred to as the baseline run in this work) has been the default setup for describing the photospheric and coronal magnetic field properties thus far. However, as shown in this investigation, results may vary more or less dramatically depending on the choice of input conditions. Before discussing trends and results for individual events and/or quantities, we remark that the approach we took in this work is that of a ``semi-hindcast'': While CME input parameters that are well-constrained by observations (such as the source region location or the CME speed and size in the corona) were derived directly from remote-sensing or in-situ data, other less-constrained parameters (such as the drag coefficient or the adiabatic index) were fine-tuned to match the MFR arrival time in the baseline run. We then ran, for each of the four events under study, 64 OSPREI simulations (16 different PFSS source surfaces for each of the four magnetograms) using a fixed set of CME and background solar wind input parameters. This allowed us to focus on ensemble variations due uniquely to the choices of input magnetogram and PFSS source surface height, rather than on the intrinsic success of a given forecast.

The four events investigated in this study were selected to originate from different locations on the Sun with respect to Earth's viewpoint---roughly, front side, west limb, east limb, and far side---and to range from the deep Solar Cycle 24/25 minimum through the ascending phase of Solar Cycle 25. Albeit far from being a comprehensive, statistical study, the results shown here can be contextualized with respect to the CME source region location and the representative global magnetic field configurations at different phases of the solar cycle. For example, the case studied here that showed the greatest dependency on the initial conditions is Event~1, for which 10 out of the 64 ensemble runs resulted in the MFR completely missing PSP. This may seem surprising given that this CME was front-sided and originated from close to the central meridian as seen from Earth (thus featuring the most up-to-date magnetograms); however, this event took place during the deep solar minimum, and its source region had no active regions in its vicinity (see Fig.~\ref{fig:magnetograms_event1}). As a result, there are no strong magnetic forces that are able to ``channel'' the CME as it propagates through the solar corona \citep[e.g.,][]{shen2011}. In such cases, the polar field strengths and flux distributions---which dictate the overall amount of closed and open magnetic flux and can significantly impact the size and shape of the helmet streamer belt---are likely to be the dominant factor for determining the large-scale gradients in magnetic pressure responsible for CME deflections \citep[see, e.g.,][for a comparison of different magnetogram sources]{riley2014, linker2017, li2021}. This is likely reflected in Event~1's estimated ForeCAT deflections being distributed over ${\sim}25^{\circ}$ in latitude and ${\sim}20^{\circ}$ in longitude. 

Another case that displayed relatively major differences is Event~3, which originated from close to the eastern limb as seen from Earth and took place ${\sim}1.5$~years after Event~1---as Solar Cycle 25 was ascending, which is also clear from the presence of multiple active regions on the photosphere (see Fig.~\ref{fig:magnetograms_event3}). This CME also displayed a significant spread in ForeCAT deflections (even more dramatically than Event~1), but the most striking result was related to the MFR tilt angle, which showed differences up to ${\sim}60^{\circ}$ at 20~R$_{\odot}$ across the ensemble. As a result, the runs that feature a large counterclockwise rotation of the MFR are those that better captured the large-scale magnetic configuration of the CME observed by PSP in FIDO---due to the variety of ForeCAT rotations observed in this case, Event~3 is the only event in our sample for which a fundamentally different magnetic configuration of the CME would be predicted depending on the photospheric and coronal input conditions. Events~2 (western limb as seen from Earth) and 4 (far side as seen from Earth), on the other hand, were characterized by the least variance in ForeCAT results and by basically consistent FIDO predictions across all the ensemble runs. Interestingly, for both these events the CME source active region is entirely ``missing'' in two of the four magnetograms (see Figs.~\ref{fig:magnetograms_event2} and \ref{fig:magnetograms_event4})---i.e., HMI Synchronic and GONG ADAPT, which are released in real time and thus contain no information on ``future'' Carrington longitudes. Given the high level of agreement across the different OSPREI results, this suggests that the magnetic environment surrounding a CME's source region may play a more important role than the source region itself in terms of coronal deflections and rotations. In fact, for Event~3 the CME source region itself is present in all the photospheric maps, but the surrounding environment is fundamentally different across different magnetogram realizations.

\begin{figure}[t!]
\centering 
 \includegraphics[width=.99\linewidth]{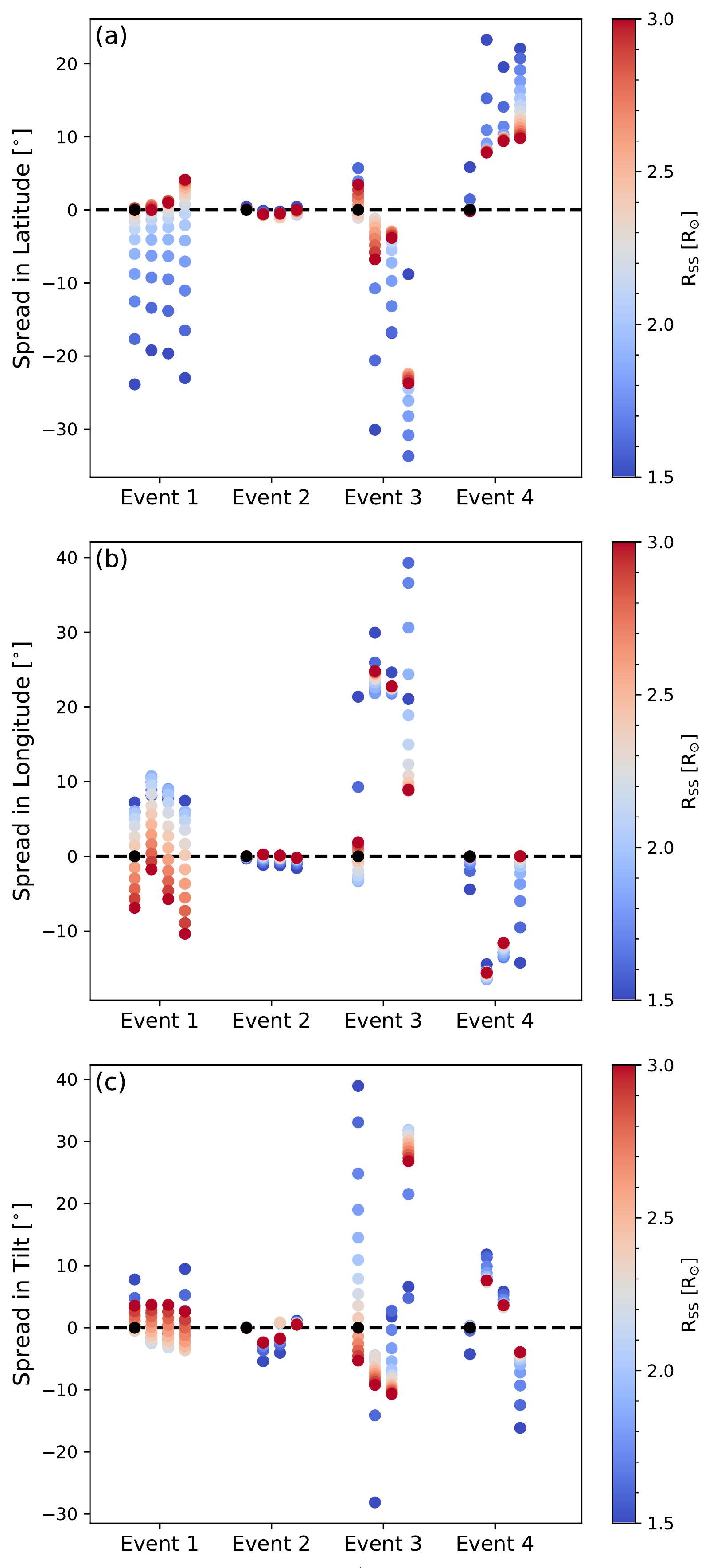}\\
  \caption{ForeCAT spreads at 20~R$_{\odot}$ in (a) latitude, (b) longitude, and (c) tilt. Events are listed on the x-axis with four columns for each event corresponding to the four magnetograms used, from left to right: HMI Synchronic, HMI Synoptic, GONG Zero, and GONG ADAPT. The columns are color-coded according to the PFSS source surface height and the deflection/rotation values are normalized according to the OSPREI ``standard'' setup of HMI Synchronic with R$\mathrm{_{ss}}$=2.5~R$_{\odot}$ for each event (represented as a black scatter point).}
  \label{fig:forecat_spreads}
\end{figure}

A visualization of the differences in ForeCAT results across the different runs for each of the events studied here is shown in Figure~\ref{fig:forecat_spreads}, which displays the spreads in latitude, longitude, and tilt at 20\,R$_{\odot}$ with respect to the baseline run---HMI Synchronic with R$\mathrm{_{ss}}=2.5$\,R$_{\odot}$. It is clear from Figure~\ref{fig:forecat_spreads} that the influence of the input magnetogram and PFSS source surface height on coronal deflections and rotations varies from case to case, with Event~3 showing significant spreads and Event~2 featuring virtually no differences with respect to the baseline run. In the case of Event~1, the distributions in latitude and longitude appear organized by the value of R$_\mathrm{ss}$, with lower (higher) source surface heights leading to the CME being directed more southward (northward) and westward (eastward). Given that this event took place during deep solar minimum conditions, in the absence of strong active region fields the location and height of the HCS cusp are likely to play a major role in channeling a CME after its eruption. In Events~3 and 4, on the other hand, the spreads appear generally less organized, possibly due to the much more complex configuration of the photospheric magnetic field during the ascending phase of the solar cycle. Overall, individual patterns in deflections and rotations may emerge based on the balance of local magnetic gradients due to active region fields and global magnetic gradients pointing to the HCS. However, the lower the R$_\mathrm{ss}$ value, the less time a CME spends under the influence of such gradients, possibly leading to reduced deflections and rotations with respect to higher R$_\mathrm{ss}$ trajectories.

Regarding FIDO results, it is clear that the input photospheric and coronal conditions have a much more substantial impact on the modeled magnetic field parameters rather than the plasma quantities---in particular, proton density and temperature appear unchanged across the ensembles, while minor differences can be observed for the solar wind speed (with Event~3 only displaying a larger spread of results). This is to be expected, since the choice of magnetogram and R$_\mathrm{ss}$ only affects the location/orientation of the CME and not its internal properties or bulk speed. Alongside differences and similarities across different input magnetograms, it is clear the choice of PFSS source surface height may affect the MFR structure predicted by OSPREI, with some events showing large distributions in results with varying R$_\mathrm{ss}$ (e.g., FIDO output of Events~1 and 3) and some showing almost no change (e.g., FIDO output for Events~2 and 4). We also note that the CME arrival time is mostly unaffected by the choice of photospheric and coronal magnetic fields, the only differences (resulting in spreads of a couple of hours at most) being related to which part of the MFR is being crossed by the spacecraft. This is also to be expected, since we used in this work a version of OSPREI that includes a simple 1D, time-independent empirical model for the ambient solar wind. Future developments to OSPREI that allow for a time-dependent, variable ambient wind \citep[see][for initial reports on the inclusion of any 1D profile for the background wind]{kay2022b} will likely result in the input photospheric and coronal conditions having an effect also on the CME propagation module of the model, i.e.\ ANTEATR. Additionally, the inclusion of solar wind high-speed streams in OSPREI will also allow for better estimates of the CME passage time at a spacecraft---in most of the events modeled here (i.e., all but Event~1), the MFR was followed by the fast solar wind, resulting in compression and/or inhibited expansion and, thus, in overestimation of the CME duration by OSPREI.

Finally, it is worth considering the results presented in this work in the context of operational CME models in space weather forecasting offices. The events we selected represent a few unique scenarios for forecasters: Event~1 is a stealth CME, Event~2 is a limb eruption as seen by SDO, Event~3 is close to the limb but seen by both SDO and STEREO-A, and Event~4 is completely on the Sun's far side. Analyzing case studies from a variety of sources with respect to Earth's viewpoint (while assuming that observations of the photospheric magnetic field are only available from Earth) is extremely beneficial for future CME predictions across the whole heliosphere \citep[e.g.,][]{shiota2014, palmerio2022a}---considering, e.g., the increasing interest in space weather at Mars \citep[e.g.,][]{lee2017, luhmann2017, palmerio2022b}, it is not unreasonable to expect that forecasts over the next decade(s) will include information for other planets and/or for crews on deep-space travel. As we have shown in this study, in the OSPREI model, the type of magnetogram and PFSS source surface height the end-user chooses will affect both the coronal deflection/rotation and MFR prediction of the CME in a more or less significant way. We do not report what combination of input conditions results in the most accurate prediction of the MFR configuration---this will be detailed in a follow-on study. In fact, in future studies, we hope to statistically determine the best input magnetogram and PFSS source surface height for a variety of CME case scenarios. For example, while this study only focuses on four events, in a larger study, many CMEs spread across a few different scenarios (e.g., far side versus on-disk eruptions, as well as solar minimum versus solar maximum ones) may provide enough results to draw correlations between properties such as the PFSS source surface height and the MFR magnetic field magnitude as modeled by OSPREI. Additionally, another future research direction could be to extend the ANTEATR module of OSPREI to include other coherent but specifically non-MFR types of CME ejecta profiles for events that may not be well-described by an MFR topology \citep{alhaddad2011, alhaddad2019a, alhaddad2019b}. In this study, our ``semi-hindcast'' approach to fine-tuning OSPREI to each CME allowed us to easily compare the model results to PSP magnetic field and plasma data. In operational use cases, a forecaster does not have readily available in-situ data of the CME to fine-tune OSPREI's input parameters, hence in future studies we will test OSPREI using only remote-sensing observations available in real time and default input parameters. Multi-spacecraft validations and comparisons to modeled-versus-observed space weather responses at different locations will also be a valuable addition to future OSPREI investigations. In fact, recent analyses of MHD simulation data have shown that internal CME structures can yield substantially different in-situ profiles depending on the observational sampling trajectories and radial distances \citep[e.g.,][]{scolini2021, lynch2022}, thus highlighting the need for multi-spacecraft observations that can provide the larger-scale heliospheric CME context and/or additional modeling constraints.


\section{Summary and Conclusions} \label{sec:conclude}

In this work, we have explored the sensitivity of the OSPREI CME analytical model to the input photospheric and coronal conditions. We have considered four events that were observed in situ by PSP, with source regions located at variable positions with respect to Earth's viewpoint. We have used four different magnetogram maps and 16 PFSS source surface heights to realize 64 ensemble runs for each of the events while keeping CME and background solar wind input parameters fixed. We found that the influence of the input photospheric and coronal fields on the ``final'' CME magnetic field predictions tends to vary from event to event: For Event~1, a subset of simulation runs resulted in the CME missing PSP altogether, Event~2 and Event~4 showed generally less disagreement between predictions, and for Event~3 different combinations of input photospheric and coronal conditions led to different MFR configuration estimates in situ. 

We found no overall pattern in the way the chosen magnetogram or PFSS source surface height affect the output of ForeCAT or FIDO, but we suggest that results tend to vary more or less dramatically with respect to the input conditions depending on the phase of the solar cycle (affecting the presence of active regions on the photosphere) and the CME source region location with respect to Earth's viewpoint (affecting how ``old'' its photospheric observations are). Our study showed that OSPREI is moderately sensitive to the input magnetogram and PFSS model used to reconstruct the photospheric and coronal magnetic environment from which the CME propagates. This realization, as well as follow-up work, will help pinpoint possible limitations of the model, but more importantly, will establish context for space weather forecasters looking to use architectures such as OSPREI---thus predicting the MFR configuration of CMEs---operationally.


\begin{acknowledgements}
This work was supported by NASA's Parker Solar Probe Guest Investigator (PSP-GI) program (grant no.\ 80NSSC22K0349).
E.P.\ also acknowledges support from NASA's HTMS program (grant no.\ 80NSSC20K1274).
N.A.\ acknowledges support from NSF AGS-1954983 and NASA ECIP 80NSSC21K0463.
P.R.\ acknowledges support from NASA's HSO-C program (80NSSC20K1285) and R2O program (80NSSC20K1403).
Solar disk and coronagraph imagery related to the events studied in this work has been inspected and analyzed via the JHelioviewer \citep{mueller2017} software.
\end{acknowledgements}

\bibliographystyle{aa}
\bibliography{master}




\end{document}